\documentclass[12pt]{article}
\usepackage[colorlinks,linkcolor=blue,citecolor=blue,urlcolor=blue,bookmarks,bookmarksnumbered]{hyperref}
\usepackage[paper=letterpaper,margin=1.0in]{geometry}
\usepackage{graphicx,xcolor}

\usepackage{amsfonts}
\usepackage{amsmath,amssymb,amsthm,bm,multirow,longtable}
\usepackage{epsfig}
\usepackage{epstopdf}
\usepackage{color}
\usepackage{cite}

\newcommand{\be}{\begin{equation}}
\newcommand{\ee}{\end{equation}}
\newcommand{\bea}{\begin{eqnarray}}
\newcommand{\eea}{\end{eqnarray}}
\newcommand{\ba}{\begin{array}}
\newcommand{\ea}{\end{array}}
\newcommand{\ben}{\begin{enumerate}}
\newcommand{\een}{\end{enumerate}}
\newcommand{\bi}{\begin{itemize}}
\newcommand{\ei}{\end{itemize}}
\newcommand{\bc}{\begin{center}}
\newcommand{\bfig}{\begin{figure}}
\newcommand{\efig}{\end{figure}}
\newcommand{\bq}{\begin{quotation}}
\newcommand{\eq}{\end{quotation}}
\newcommand{\bt}{\begin{table}}
\newcommand{\et}{\end{table}}
\newcommand{\btab}{\begin{tabular}}
\newcommand{\etab}{\end{tabular}}
\newcommand{\bs}{\begin{slide}}
\newcommand{\es}{\end{slide}}

\newcommand{\pa}{\partial}

\newcommand{\IR}{\mathbb{R}}

\newcommand{\X}{\mathbb{X}}

\newcommand{\ket}{\rangle}

\def\pa{\partial}
\def\om{\omega}

\def\S{\mathbb{S}}
\def\s{\sigma}

\def\ket{\rangle}

\def\S{\mathbb{S}}

\def\ket{\rangle}

\def\S{\mathbb{S}}

\newcommand{\K}{\mathbb{K}}
\newcommand{\comment}[1]{}

\def\pa{\partial}
\def\om{\omega}

\def\S{\mathbb{S}}
\def\s{\sigma}

\def\Ph{\cal{P}}

\def\ket{\rangle}

\def\S{\mathbb{S}}
\def\s{\sigma}

\def\Ph{{\cal{P}}}

\def\ket{\rangle}

\def\tx{\tilde{x}}

\newcommand{\Z}{\mathbb{Z}}
\newcommand{\R}{\mathbb{R}}
\def\tk{\tilde{k}}

\newcommand{\tR}{{\tilde R}}

\newcommand{\beq}{\begin{eqnarray}}
\newcommand{\eeq}{\end{eqnarray}}

\let\ba=\overline

\def\e{\epsilon}

\let\l=\lambda
\let\L=\Lambda

\def\tx{\mathord{\tilde x}}
\def\htx{\mathord{\hat{\tilde x}}}

\let\w=\omega

\def\IR{\relax\leavevmode{\rm I\kern-.18em R}}
\def\ZZ{\relax\leavevmode
       \ifmmode\mathchoice
       {\hbox{\sf Z\kern-.4em Z}}
       {\hbox{\sf Z\kern-.4em Z}}
       {\lower.9pt\hbox{\scriptsize\sf Z\kern-.36em Z}}
       {\lower1.2pt\hbox{\tiny\sf Z\kern-.36em Z}}
       \else{\sf Z\kern-.4em Z}\fi}
\def\RR{\relax\leavevmode
       \ifmmode\mathchoice
       {\hbox{\sf R\kern-.4em R}}
       {\hbox{\sf R\kern-.4em R}}
       {\lower.9pt\hbox{\scriptsize\sf R\kern-.36em R}}
       {\lower1.2pt\hbox{\tiny\sf R\kern-.36em R}}
       \else{\sf R\kern-.4em R}\fi}

\def\resetby#1#2{\@addtoreset{#2}{#1}}
\def\seceq{\@addtoreset{equation}{section}
              \def\theequation{\thesection.\arabic{equation}}}

\def\Label#1{\label{#1}%
                \smash{\hbox to0pt{\raise1ex\hbox{\tiny[#1]}\hss}}}
\def\noLabels{\let\Label=\label}

\def\TeV{\text{T\kern0pte\kern-1ptV}}
\begin{document}
\thispagestyle{empty}

\vskip 0.5cm
\title{\bf From Quantum Foundations of Quantum Field Theory, String Theory and Quantum Gravity to Dark Matter and Dark Energy}
\author{\normalsize
        %\textsuperscript1,
        %\textsuperscript2 and
        {\large \bf Djordje Minic}%\textsuperscript1
\\*
        %\footnotesize\textsuperscript1\it\,\\*[-5pt]
        %\footnotesize\textsuperscript2\it\,\\*[-5pt]
        %\footnotesize\textsuperscript1\it\,
{\large \it Department of Physics, Virginia Tech, Blacksburg, VA 24061, U.S.A.}\\
                                                        dminic@vt.edu} 
\date{\small\today}

\maketitle
\begin{abstract}\noindent
We review our recent  work on quantum foundations of quantum mechanics, quantum field theory
and quantum gravity (formulated as metastring theory) and various implications for the problems
of dark matter and dark energy. The first point concerns the new understanding of quantum theory
via the concept of quantum (modular) spacetime endowed with manifest non-locality that is consistent with causality.
This view implies the consistency of the fundamental length and Lorentz symmetry, based on the principle
of relative (observer dependent) locality. The geometry of such quantum spacetime is encoded in the new concept
of Born geometry. This in turn leads to a novel understanding of quantum field theory in a manifestly bi-local 
representation endowed with metaparticle quanta. A fully dynamical quantum spacetime, with a dynamical
Born geometry, leads to quantum gravity (a quantum theory of dynamical geometry of quantum non-locality) 
in the guise of metastring theory.
This generic formulation of string theory implies a radiatively stable positive cosmological constant (viewed as a
curvature of the dual spacetime) as a model of dark energy in
the observed classical spacetime, as well as metaparticle quanta (the zero modes of the metastring) 
as the natural quanta of dark matter in this approach.
\end{abstract}

\section{Introduction and Overview}

In this talk I 
review recent work 
\cite{Freidel:2013zga, Freidel:2014qna, Freidel:2015pka, Freidel:2015uug, Freidel:2016pls, Freidel:2017xsi, Freidel:2017wst, Freidel:2017nhg, Freidel:2018apz, Freidel:2019jor}
on quantum foundations of quantum mechanics, quantum field theory
and quantum gravity (in the form of metastring theory) as well as unique implications for the problems
of dark matter and dark energy. The starting point addresses the new understanding of quantum theory
using the concept of quantum, or modular, spacetime endowed with manifest non-locality that is consistent with causality.
This view implies the consistency of the fundamental length and Lorentz symmetry, based on the principle
of observer based, or relative, locality. The geometry of such quantum spacetime is encoded in the new concept
of Born geometry. This leads to a new understanding of quantum field theory in a manifestly bi-local 
representation endowed with metaparticle quanta. A fully dynamical quantum spacetime (a dynamical
Born geometry) leads to a theory of quantum gravity in the form of
metastring theory (a robust quantum theory whose geometry is the Born geometry of quantum non-locality).
This generic formulation of string theory implies a radiatively stable positive cosmological constant 
(dark energy) \cite{rBHM8} in
the observed classical spacetime and metaparticle quanta (the zero modes of the metastring) 
representing the natural quanta of dark matter \cite{Ho:2010ca} (correlated to dark energy
and visible matter).

The logic of our story is very similar to the path that leads from the Minkowski geometry of special relativity via relativistic non-gravitational field theory
to a dynamical spacetime of general relativity.
In our case we start with a hidden geometry in quantum theory (Born geometry) 
and proceed to its dynamical implementation
in quantum gravity (formulated as a metastring theory) with implications for quantum field theory 
(formulated in a way that takes into account the hidden Born geometry) with implications for the
observed world: metaparticles as dark matter quanta, and dark energy emerging from the geometry of
the dual spacetime.
In some sense this story is a sharpening of the modern approaches to non-perturbative quantum physics 
\cite{Wilson:1973jj}, using a simple but crucial insight about a completeness of quantum kinematics 
of discretized physical systems \cite{Zak:1968zz}.
%%%%%%%%%%%%%%%%%%%%%%%%%%%%%%%%%%%%%%%%%%%%%%%%%
%%%%%%%%%%%%%%%%%%%%%%%%%%%%%%%
%%%%%%%%%%%%%%%%%%%%%%%%%%%%%%%%%

In particular, in \cite{Freidel:2016pls} we have shown that any quantum theory is endowed with
a generic quantum polarization associated with modular spacetime \cite{Freidel:2015uug}.
The generic polarization manifestly realizes quantum non-locality (associated with quantum superposition principle)
that is consistent with causality, and reveals a novel underlying geometry structure, that we call Born geometry 
\cite{Freidel:2013zga, Freidel:2014qna}, which unifies symplectic, orthogonal and conformal geometries.
Born geometry turns out to be fundamental in a particular quantum theory that consistently propagates
in this geometry - this turns out to be string theory formulated in a generalized-geometric and intrinsically non-commutative,
doubled, form (that we call metastring theory) \cite{Freidel:2015pka,  Freidel:2017xsi, Freidel:2017wst, Freidel:2017nhg}.
The zero modes of the metastring define a new concept, called metaparticle, that explicitly realizes the
geometry of modular spacetime \cite{Freidel:2018apz}, and that could be considered as an explicit
prediction of the modular representation of quantum theory. The metaparticle \cite{Freidel:2019jor} is associated with a modular
generalization of quantum fields which can be viewed as low energy remnants of metastring fields.

Note that from this new viewpoint \cite{Freidel:2013zga, Freidel:2014qna, Freidel:2015pka, Freidel:2015uug, Freidel:2016pls, Freidel:2017xsi, Freidel:2017wst, Freidel:2017nhg, Freidel:2018apz, Freidel:2019jor} 
quantum gravity is essentially defined as ``gravitization of the quantum'', that is, as a theory of
a dynamical Born geometry (for more on Born geometry, see \cite{Freidel:2017yuv}). As such it
incorporates the concept of Born reciprocity \cite{born} as a covariant implementation of T-duality, the fundamental
relation between short and long distance physics in string theory, as well as the new idea of relative (or observed dependent) locality  \cite{AmelinoCamelia:2011bm}.

In the first part of this talk we describe the hidden quantum spacetime geometry underlying the generic representation of quantum theory,
including quantum field theory (which renders the generic modular representation manifestly non-local)
and then in the second part we find that the same (and, in general, dynamical) geometric structure
underlies metastring theory, a manifestly T-duality covariant formulation of string theory, viewed as a consistent theory 
of quantum gravity and matter. Thus quantum gravity ``gravitizes'' the quantum spacetime geometry.
Finally, in the last part we discuss a robust effective description of such a theory of quantum gravity at long distance that leads to a non-commutative (yet covariant) effective field
theory, with metaparticle quanta, as implied by an intrinsic non-commutativity of closed string theory, and we discuss some of its natural consequences for
the problems of dark energy and dark matter.

\section{Quantum theory from quantum spacetime}

The fundamental reason for the existence of modular polarizations in quantum theory 
can be seen as follows \cite{Freidel:2016pls}:
If one imagine that a quantum system is formulated on a lattice (as assumed in the
modern (Wilsonian) non-perturbative approaches \cite{Wilson:1973jj}), then a theorem due to Zak \cite{Zak:1968zz}
states that a complete set of quantum numbers needed to describe any quantum system would require both
quantum numbers associated with the lattice and its inverse. 
This is easy to see by realizing that non-commuting Hermitian operators, such as coordinates and momenta,
$[q, p] = i \hbar$, when exponentiated, together with the appropriate lattice spacing $a$ and 
its inverse $\frac{2\pi \hbar}{a}$,
commute, that is, $[\exp(\frac{i}{\hbar} q \frac{2 \pi \hbar}{a}), \exp(\frac{i}{\hbar} p a)]=0$.
Such unitary observables were labeled as ``modular'' by Aharonov \cite{Aharonov:1969qx}, and have also
appeared in other discussions of the fundamental issues in quantum theory \cite{schwinger}. 
Note that these variables are purely quantum, in the sense that their formal $\hbar \to 0$ limit is singular.
Also, even their commutators are zero, the associated Poisson brackets are non-zero, as these are unitary (phase)
variables. Finally, the classical limit is defined by starting with a modular formulation and defining an
appropriate ``extensification" \cite{Freidel:2016pls}.
Thus, in principle, there are many consistent classical limits (as suggested by 
the consistent history approach to quantum theory \cite{GellMann:1992kh}).
As we shall in what follows these purely quantum variables will appear in the context of 
quantum field theory as well.

So we are instructed to look at a complete set of unitary operators (as opposed to a complete
set of Hermitian operators). Let us look at the simplest example of the 
$q$ and $p$ operators. The commuting subalgebra of the original non-commuting $[q, p] = i \hbar$ algebra, can be completely described by self-dual lattices (endowed with the natural symplectic 
form ($\omega$) coming from the commutator bracket). These in turn represent a discretization of a phase space
(in general fully covariant) defined by $q$ and $p$ and when lifted to
the original non-commutative algebra, require extra data associated with the lift that is described by
a doubly orthogonal ($O(d,d)$, where $d$ denotes the spacetime dimension) 
metric $\eta$ (a symmetric counterpart of the antisymmetric $\omega$, associated with $Sp(2d)$ transformations).
Finally, in order to define the vacuum state on this self-dual lattice, we need a conformal structure
$O(2, 2(d-1))$ \cite{Freidel:2016pls}. This triplet of structures define what we call Born geometry 
\cite{Freidel:2013zga, Freidel:2014qna} associated with the modular representation of quantum theory
\cite{Freidel:2016pls}, which naturally captures quantum non-locality that is consistent with causality, given
the quantum nature of the unitary operators and the fact that the triple intersection of $Sp(2d)$, $O(d,d)$
and $O(2, 2(d-1))$, gives the Lorentz group \cite{Freidel:2017xsi}.

We now formalize these insights about the hidden quantum spacetime geometry of quantization \cite{Freidel:2016pls}, which will, 
perhaps surprisingly, take us all the way to quantum gravity in the guise of an intrinsically non-commutative formulation of string theory.
We start with the Heisenberg (or Weyl-Heisenberg) group, which is generated, on the level of the corresponding
algebra, by the familiar position $\hat{q}^a$ and momentum $\hat{p}_b$ operators:
\begin{equation}
[ \hat{q}^a, \hat{p}_b] = i \hbar \delta^a_b .
\end{equation}
It will be convenient to introduce a length scale $\lambda$ and a momentum scale $\epsilon$, 
with $\lambda \epsilon = \hbar$.
Then, let us introduce the following notation
$
\hat{x}^a \equiv \hat{q}^a/\lambda, \quad \hat{\tilde{x}}_a \equiv \hat{p}_a/\epsilon, 
$, with $ [ \hat{x}^a, \hat{\tilde{x}}_b] = i \delta^a_b .
$
Even more compactly let us suggestively write 
\begin{equation}
\X^{A}\equiv (x^a , \tilde{x}_a)^{T}, \quad [ \hat{\X}^a, \hat{\X}^b] = i \omega^{AB} ,
\end{equation}
with $\frac{1}{2}\omega_{AB} d X^A d X^B = \frac{1}{\hbar} dp_{a} \wedge dq^{a}$, where 
$\omega_{AB} = - \omega_{BA}$ is the canonical symplectic form on phase space $\cal{P}$.
The Heisenberg group $H_{\cal{P}}$ is generated by Weyl operators \cite{weyl}
($\K$ stands for the pair $(\tk,k)$ and  $\omega(\K,\K')=k \cdot \tilde{k}' - \tilde{k}\cdot k'$)
\begin{equation}
W_{\K} \equiv e^{2 \pi i \omega( \K, \X)}.
\end{equation}
These form a central extension of the translation algebra
\begin{equation}
W_{\K}W_{\K'} = e^{2 \pi i \omega( \K, \K' )} W_{\K + \K'}.
\end{equation}
The projection $\pi : H_{\cal{P}} \to \cal{P}$ (where $\pi : W_{\K} \to \K$) defines a line bundle over $\cal{P}$ (in principle a covariant phase space of quantum probes).
In this formulation, states are sections of degree one 
\begin{equation}
W_{\K'} \Phi(\K) = e^{2 \pi i \omega( \K, \K' )} \Phi(\K + \K').
\end{equation}
In this language, geometric quantization means to take a Lagrangian $L \in \cal{P}$, so that states descend to
square integrable functions on $L$.

A Lagrangian submanifold $L$ is a maximally isotropic subspace $L$ with $\omega|_L =0$,
and thus $\{ \partial/\partial q^a \} \in T \cal{P}$ defines a Lagrangian submanifold, or ``space''.
(Indeed, $\omega(\partial/\partial q^a , \partial/\partial q^b) = 0$.)
This can be understood as a classical characterization of space (and in the covariant context, of spacetime),
as a ``slice'' of phase space. How about a purely quantum characterization of space? We claim that quantum theory reveals a new notion of quantum space (and, more covariantly, a new notion of quantum spacetime).

Note that for space-like separations the operators of a local quantum field theory commute. Thus in order
to understand the meaning of quantum spacetime (quantum Lagrangian), we need
to look at a maximally commuting subalgebra of the Heisenberg algebra and the representation that diagonalizes it.
Thus, borrowing from notions of non-commutative algebra and non-commutative geometry \cite{connes} (such as the theorem of Gelfand-Naimark \cite{GN}), we can say that a 
Lagrangian submanifold is a maximally commutative subgroup of the Heisenberg group.
If we accept this notion of a Lagrangian, then the quantum regime is very different from the
classical regime.
In particular the vanishing Poisson bracket $\{f(q), g(p) \}$ requires either $f$ or $g$ to be constant.
However, the vanishing commutator $[f(\hat{q}), g(\hat{p})]=0$ requires only that the functions be commensurately
periodic
\begin{equation}
e^{i \alpha \hat{p}} e^{i \beta \hat{q}} = e^{i \hbar \alpha \beta} e^{i \beta \hat{q}} e^{i \alpha \hat{p}} , 
\quad  \alpha \beta = 2 \pi/\hbar.
\end{equation}
What is interesting here is that similar considerations led Aharonov to introduce {\it modular variables} to describe purely quantum phenomena,
such as interference \cite{aharonov2008quantum}. 

\subsection{Modular variables}

Modular variables are described in great detail in the very insightful book by Aharonov and Rohrlich \cite{aharonov2008quantum}, where one can find detailed bibliography on this subject\footnote{See also  
\cite{schwinger, Aharonov:1969qx}.}. 
The fundamental question posed there was as
follows: how does one capture interference effects (due to the fundamental linearity of quantum theory) in terms of 
Heisenberg operators?
For example, what are the quantum observables that can measure the relative  phase responsible for interference in a double-slit experiment? No polynomial functions of the operators $\hat{q}$ and $\hat{p}$ can detect such phases, but operators that translate in space, such as $e^{i R \hat{p}/\hbar}$, do. 
Thus the modular  variables denoted $[\hat{q}]$ and $[\hat{p}]$, which are defined modulo a length scale $R$ (the slit spacing being a natural choice), play a central role, where
\begin{equation}\label{QMmodvars}
[p]_{2 \pi \hbar/R} = p \;\; \mathrm{mod}\left(2 \pi \hbar/R\right),\qquad  
[q]_R= q \;\; \mathrm{mod}\left( R\right).
\end{equation}
The shift operator $e^{i R\hat{p}/\hbar}= e^{iR [\hat{p}]/\hbar}$ shifts the position of a particle state (say an electron in the double-slit experiment) by a distance $R$ and is a function of the modular momenta\footnote{There is a very close similarity here to the Galois quantum theory discussed in \cite{Chang:2012eh}.}.  
These modular variables (the main examples being the Aharonov-Bohm and Aharonov-Casher phases \cite{aharonov2008quantum}) satisfy non-local operator equations of motion. For example, given the Hamiltonian,
$\hat{H} = \hat{p}^2/2m +V(\hat{q})$, the 
Heisenberg equation of motion for the shift operator is,
\begin{equation}
e^{-i R\hat{p}/\hbar}  \,\frac{d }{ dt}e^{iR\hat{p}/\hbar} = -   \frac{iR }{\hbar}\left(\frac{V(\hat{q}+R)-V(\hat{q})}{R}\right).
\end{equation}
Modular variables are fundamentally
 non-local in a non-classical sense, since we see here that their evolution depends on the value of the potential at distinct locations. 
Remarkably, thanks to the uncertainty principle, this dynamical non-locality does not lead to a violation of causality \cite{aharonov2008quantum}.
One of the characteristic features of these variables is that they do not have classical analogues; indeed, the limit $\hbar \to 0$ of $[p]_{h /R}$ is ill-defined. Also modular variables capture entanglement of continuous $q$, $p$ variables.

Note that modular variables are, in general, covariant and, also, contextual\footnote{Aharonov and collaborators have pushed the logic associated with modular variables to argue for a new kind of weak measurements of such non-local variables that capture the superposition principle of quantum theory. Similarly, Aharonov and collaborators argue for a time symmetric formulation of quantum theory \cite{aharonov2008quantum}.}. In other words, they carry specific experimental information, such as the length $R$ between
the two-slits. However, in the context of quantum gravity such scales are automatically built in, and the contextuality is in principle removed. Also, the fundamental dynamical equations for modular variables are non-local in quantum gravity because of the presence of the fundamental length.

When exponentiated (i.e. when understood as particular Weyl operators), the modular variables naturally commute. In other words, given $[x^a, \tilde{x}_b] = \frac{i}{2 \pi} \delta^a_b$, the following commutator of modular operators vanishes \cite{Freidel:2016pls}
\begin{equation}
[e^{2\pi i x}, e^{2\pi i \tilde{x}}] =0.
\end{equation}
Thus a quantum algebra of modular variables possesses more commutative directions than a classical Poisson algebra,
because the Poisson bracket of modular variables does not vanish, $\{ e^{2\pi i x}, e^{2\pi i \tilde{x}} \}\neq 0$.

Here we make a historical note \cite{history}: The above non-local equations of motion were essentially written by Max Born, in the very first paper which used the phrase ``Quantum Mechanics'' in its title, in 1924, one year before the Heisenberg breakthrough paper.
Actually, Heisenberg crucially used Born's prescription of replacing classical equations by the corresponding difference equations, in order to derive what we now call the canonical commutation relations (properly written by Born and Jordan) from the Bohr-Sommerfeld quantization conditions.\footnote{{Note that the above discrete operatorial (Born-Heisenberg-Jordan) equations of motion for modular variables can be generalized to
other quantum systems \cite{Adler:1988hb}. Also, the appearance of the ``covariant hidden spacetime $\tilde{x}$'' bears some similarity to the
Koopman-von-Neumann theory \cite{KvN}.}}

\subsection{Modular spacetime  and Born geometry of quantum theory}

Returning to the subject of quantum Lagrangians, notice that the quantum Lagrangian is analogous to a Brillouin cell in condensed matter physics.
The volume and shape of the cell are given by $\lambda$ and $\epsilon$ (i.e. $\hbar$ and $G_N$ ($\alpha'$))
The uncertainty principle  is implemented in a subtle way: we can specify a point in modular cell, but if so, we can't say {\it which} cell we are in.

This means that there is a more general notion of quantization \cite{Freidel:2016pls}, beyond that of geometric quantization.
Instead of selecting a classical polarization $L$ (the arguments of the wave function, or the arguments of a local quantum field)
we can choose a {\it modular polarization}.
In terms of the Heisenberg group all that is happening is that in order to have a commutative algebra, we need only
\begin{equation}
\omega(\K, \K') \in 2 Z, \quad W_{\K}W_{\K'} = e^{2 \pi i \omega( \K, \K' )} W_{\K + \K'} = W_{\K'}W_{\K'}.
\end{equation}
This defines a lattice $\Lambda$ in phase space $\cal{P}$.
Finally, we specify a ``lift'' of the lattice from the phase space $\cal{P}$ to the Heisenberg group $H_{\cal{P}}$.

Maximally commuting subgroups $\hat{\Lambda}$ of the Heisenberg group correspond to lattices that are integral and self-dual
with respect to $\omega$ \cite{Mackey}. 
Given $W_{\lambda}$ where $\lambda \in \Lambda$ there is a lift to $\hat{\Lambda}$  which defines ``modular polarization''
\begin{equation}
U_{\lambda} = \alpha(\lambda) W_{\lambda},
\end{equation}
where $\alpha(\lambda)$ satisfies the co-cycle condition
\begin{equation}
 \alpha(\lambda) \alpha(\mu) e^{ \pi i \omega( \lambda, \mu)} =\alpha(\lambda + \mu), \quad \lambda, \alpha \in \Lambda.
\end{equation}
One can parametrize a solution to the co-cycle condition by introducing a symmetric bilinear from $\eta$ and setting
(with $\eta(\K,\K') =k \cdot \tilde{k}' + \tilde{k}\cdot k',$)
\begin{equation}
 \alpha_{\eta}(\lambda) \equiv e^{i \frac{\pi}{2}  \eta( \lambda, \lambda)} .
\end{equation}
Finally, when we choose a classical Lagrangian $L$, there is a special state that we associate with the vacuum:
it is translation invariant (which in our context can be interpreted as ``empty space'').
In modular quantization there is no such translation invariant state (because of the lattice structure).
The best we can do is to choose a state that minimizes an ``energy'', which requires the introduction of another
symmetric bilinear form, that we call, again suggestively, $H$.
This means, first, that we are looking for 
operators such that 
\be\label{Comm1}
 [\hat{\mathbb{P}}_A, \Phi]= \frac{i}{2\pi}\partial_A \Phi ,\qquad \Phi(\hat{\X}+\lambda)=\Phi(\hat\X),
 \ee
where the modular observables  $\Phi(\hat{\X}+{\lambda})=\Phi(\X)$ are  generated by the lattice observables $U_\lambda$ with $\lambda \in \Lambda$. 
Translation invariance would be the condition $\hat{\mathbb{P}}|0\ket=0$. Since this is not possible, the next natural choice is to minimize the translational energy. 
Therefore we pick a positive definite metric $H_{AB}$ on $\Ph$, and we define \cite{Freidel:2016pls}
\be
\hat{E}_H\equiv H^{AB} \hat{\mathbb{P}}_A\hat{\mathbb{P}}_B,
\ee  
and demand that $|0\ket_H$ be the ground state of $\hat{E}_H$.
This is indeed the most natural choice and it shows that we cannot fully disentangle kinematics (i.e., the definition of translation generators) from dynamics.
In the Schr\"odinger case, since the translation generators commute, the vacuum state $\hat{E}|0\ket=0$ is also the translation invariant state and it carries no memory of the metric $H$ needed to define the energy.
In our context, due to the non-commutativity of translations, the operators $\hat{E}_H$ and $\hat{E}_{H'}$ do not commute. As a result the vacuum state depends on $H$, in other words $|0\ket_H\neq |0\ket_{H'}$, and it also possesses a non-vanishing zero point energy.

Thus, modular quantization involves the introduction of three quadratic forms $(\omega, \eta, H)$, i.e. what we call {\it Born geometry} \cite{Freidel:2013zga, Freidel:2014qna}, which underlies the geometry of modular variables.

As we will see, in the context of metastring theory, a choice of polarization is a choice of a spacetime within
$\cal{P}$ but the most general choice is a {\it modular polarization} that we have discussed above.
From the foundational quantum viewpoint Born geometry $(\omega, \eta, H)$ arises as a parametrization of such quantizations, which
results in a notion of quantum spacetime, that we call {\it modular spacetime}. {In particular, a one dimensional (1d) modular line is a two dimensional (2d) torus that is compact and not-simply connected.}
Finally, large spacetimes  of canonical general relativity (and its extensions, like string theory) result as a ``many-body'' phenomenon, through a process of tensoring (entanglement) of unit modular cells, that we refer to as ``extensification'' \cite{Freidel:2016pls}.

In particular, the symplectic structure $\omega$ found in
$
ds^2_{\omega}= \frac{1}{2}\omega_{AB} d \X^A d \X^B = \frac{1}{\hbar} dp_{a} \wedge dq^{a},
$
is encoded in the canonical Heisenberg commutator between $q^{a}$ and $p_{a}$.
The generalized, quantum, metric $H$ comes from the Born rule in quantum theory
$
ds^2_{H} = H_{AB} d \X^A d \X^B = \frac{1}{\hbar} (\frac{dq_{a} dq^{a}}{G_N} + G_N {dp_{a} dp^{a}}).
$
For weak gravity, this metric reduces to the spacetime metric (where spacetime can be viewed as a slice of phase space).
Due to gravity's extreme weakness, we only see spacetime metric at low energies. (The ratio $\epsilon/\lambda$ defines a tension; if this is identified with $c^3/G_N$, it is enormous, $\sim 10^{32} kg/sec$.)
Therefore, in this formulation the usual
dynamical spacetime metric is the low energy leftover of the quantum metric.
Finally, the polarization  (or locality metric) $\eta$ encodes the distinction between spacetime-like and energy-momentum-like
aspects of phase space (and in this sense it defines an analog of the ``causal'' structure in phase space)
$
ds^2_{\eta} = \eta_{AB} d \X^A d \X^B = \frac{2}{\hbar}{dp_{a} dq^{a}}.
$
This new metric captures the essence of relative locality - when $\eta$ is constant we have absolute locality. Curving $\eta$ also means ``gravitizing the quantum''.
In general all three elements of Born geometry, $\omega$, $\eta$ and $H$ are dynamical and curved in metastring theory,
as we will discuss in what follows.

Also, we have that the Lorentz group (in $D$ spacetime dimensions) lies at the intersection of the symplectic, neutral and doubly orthogonal groups  \cite{Freidel:2016pls},
\be\label{orthinter}
 {\rm O}(1,d-1)={\rm Sp}(2d) \cap {\rm O}(d,d) \cap {\rm O}(2,2(d-1)),
\ee
which sheds new light on the origin of quantum theory through compatibility of the causal (Lorentz) structure and non-locality 
captured by the discreteness of quantum spacetime. This also captures the role of relative (observer-dependent) locality
\cite{AmelinoCamelia:2011bm}
needed to resolve the apparent contradiction between discreteness of quantum spacetime and Lorentz symmetry.

Let us end this discussion of quantum mechanics by a few comments regarding the Stone-von Neumann theorem \cite{S} 
which asserts that all representations of the Heisenberg group are unitarily equivalent. Normally, we think of this as a choice of basis in phase space (a choice of polarization or classical Lagrangian), and all such choices are related by Fourier transform.
Similarly, one can pass from a classical polarization (such as the  Schr\"odinger representation) to a modular polarization 
via the Zak transform \cite{Zak:1968zz} (see section 3). Note that, there is a connection on the line bundle over phase space that has unit flux through a modular cell. (This is very similar to Integer Quantum Hall effect.)
A modular wave function is quasi-periodic
\begin{equation}
\Psi(x+a, \tilde{x}) = e^{2 i \pi a \tilde{x}} \Psi (x, \tilde{x}), \quad \Psi(x, \tilde{x} + \tilde{a}) =\Psi (x, \tilde{x}).
\end{equation}
The quasi-periods correspond to the tails of an Aharonov-Bohm \cite{Aharonov:1959fk} potential attached to a unit flux.
In particular, vacuum states must have at least one zero in a cell, which leads to theta functions (the
Zak transforms of Gaussians). Note that from the point of modular polarization, the familiar  Schr\"odinger polarization is just a singular limit\footnote{{
This discussion can be also extended to a path integral formulation in modular polarization as done by Yigit Yargic in his Perimeter Scholars International master thesis work under Laurent Freidel.
}}.

\section{Quantum field theory and quantum spacetime}

Now we make some general comments about quantum field theory in the modular form, following the general modular formulation
of any quantum theory. (Later we will discuss how such a formulation of quantum field theory comes
emerges from quantum gravity at large distances.) This new modular polarization of quantum field theory reveals new structures and sheds
new light on both the short distance (UV) and long distance (IR) physics of quantum fields, and the new definition of the
continuum limit of quantum field theories which is self-dual with respect to the UV and IR properties
(resembling some crucial properties of non-commutative field theories \cite{Douglas:2001ba}. 
In particular, the modular representation of quantum field theory introduces dual ``electric'' and ``magnetic''
variables, which are in general, non-commuting. This general message extends our results in
the context of the 2d conformal field theory formulation of string theory in which the explicit non-commutativity
of such ``electric'' and ``magnetic'' variables has been explicitly demonstrated \cite{Freidel:2017wst, Freidel:2017nhg}.

The general modular representation can be defined in terms of the Zak transform of a Schr\"{o}dinger
representation (given in terms of wave functions).
Given a square normalizable wave function $\psi(x)$ (where $x\equiv q/a$ and  $\tilde{x} \equiv p/b$,
and where $ab = 2 \pi \hbar$) belonging to a Hilbert space, 
one defines the modular representation as the following lattice Fourier transform (or Zak transform)
\be
\psi_a(x, \tilde{x}) \equiv \sqrt{a} \sum_n e^{-2 \pi i n \tilde{x}} \psi(a(x +n)) .
\ee
Note that if $\psi(x)$ is a Gaussian, its Zak transform, the modular $\psi_a (x, \tilde{x})$ is given
by the doubly-period theta function associated with the lattice.
(The inverse Zak transform
\be
\Phi(x +n ) \equiv \frac{1}{\sqrt{a}} \int_0^1 d \tilde{x} e^{2 \pi i n \tilde{x}} \Phi_a (a^{-1} x, \tilde{x}) ,
\ee
illustrates that the usual Schr\"{o}dinger representation is really singular, and thus not generic.)

Now, if one second quantizes $\psi(x)$ one naturally ends up with a quantum field operator $\phi(x)$.
Similarly, the second quantization of the modular $\psi_a (x, \tilde{x})$ would lead to
a modular quantum field operator $\phi (x, \tilde{x})$
\be
\phi(x) \to \phi (x, \tilde{x}).
\ee
Note that the usual wave functional approach to quantum field theory (defined in 
terms of functionals $\Psi[\phi(x)]$ should be now
defined in terms of wave functionals of modular fields $\Psi[\phi (x, \tilde{x})]$, where the
fields $\phi$ and their duals $\tilde{\phi}$ in general do not commute, and thus the wave functional can
be still chosen to be a functional of $\phi$ in a very particular polarization. However, now we have more freedom
in the general modular polarization.
Thus the dual momenta $p$ and $\tilde{p}$ (to $x$ and $\tilde{x}$ respectively) lead, via the canonical minimal coupling prescription, not only to the
usual fields $\phi$ but also to their duals $\tilde{\phi}$ (see section 5).
This procedure defines the modular polarization of quantum field theory in terms of the
functional Zak transform of the original wave functional 
\be
\Psi[\phi(x)] \to \Psi[\phi(x, \tilde{x}), \tilde{\phi}(x, \tilde{x})] .
\ee
For example, the Gaussian wave functionals with non-trivial momentum kernels (such as the
ones found in the context of non-trivial interacting theories like 
2+1 and 3+1 dimensional compact QED \cite{Polyakov:1975rs} 
as well as 2+1 and 3+1 dimensional Yang Mills theory
\cite{Feynman:1981ss} 
would be mapped into functional theta functions.

Once again, we can understand this new view of continuum quantum field theory as 
a sharpening of the canonical picture of a quantum field theory formulated on a lattice (as assumed in the
modern, Wilsonian, non-perturbative approaches \cite{Wilson:1973jj}), and a theorem due to Zak \cite{Zak:1968zz}
which states that a complete set of quantum numbers needed to describe any quantum system would require both
quantum numbers associated with the lattice and its inverse. 
Thus, in general, before we take a naive continuum limit, we need to work with the full set of quantuum spacetime data,
represented by the lattice and its dual (spacetime and its dual) as well as fields and dual fields.

On the level of the path integral formulation (the open path integral defines the above wave functional in
terms of boundary data), one would have to work with an explicit 
formulation of quantum field theory in terms of (``electric'') quantum fields $\phi(x, \tilde{x}$) and their (``magnetic'')
duals $\tilde{\phi}(x, \tilde{x})$
\be
\int ``D\phi(x, \tilde{x}) D\tilde{\phi}(x, \tilde{x})'' e^{\frac{i}{\hbar} S_{chiral}[\phi(x, \tilde{x}), \tilde{\phi}(x, \tilde{x})  ]}.
\ee
There is no overcounting here, because the action will turn out to be chiral.
So even though one has formally doubled the number of variables, the chiral nature
of the theory, keeps the total number of degrees of freedom the same as in the
purely electric formulation.
Similarly the measure of the path integral takes into account the
chiral nature of this formally doubled formulation of quantum field theory.
That is why the doubled measure is written under the quotation marks.
By formally introducing the doublet $\Z= (\phi(x, \tilde{x}), \tilde{\phi}(x, \tilde{x}))$,
we can write this doubled path integral as
\be
\int (D \Z)_{chiral} e^{\frac{i}{\hbar} S_{chiral}[\Z]} .
\ee
As usual, this formulation has to be properly regulated, but now, with two cut-offs, with
the continuum limit defined in a symmetric, self-dual way with respect to the double RG flows
thus resembling non-commutative field theory \cite{Douglas:2001ba}, 
albeit with full spacetime covariance. This new insight on quantum field theory should be important 
not only in the context of quantum non-locality in quantum field theory, but also in the realms of
strong coupling and deep infrared.

\section{Quantum gravity and dynamical quantum spacetime}

The unexpected outcome of our research on the foundations of quantum mechanics and quantum field theory is that this fundamental quantum geometry of quantum theory can be realized in
the context of metastring theory, where this quantum geometry is ``gravitized'' (i.e. dynamical).
At the classical level, metastring theory \cite{Freidel:2013zga, Freidel:2014qna, Freidel:2015pka, Freidel:2015uug, Freidel:2016pls, Freidel:2017xsi, Freidel:2017wst, Freidel:2017nhg, Freidel:2018apz, Freidel:2019jor} can be thought of as a formulation of string theory in which the target space is doubled in such a way that T-duality acts linearly on the coordinates. This doubling means that  momentum and winding modes appear on an equal footing.  We refer to the target space as a phase space since the metastring action requires the presence of a background symplectic form $\omega$. The metastring formulation also requires the presence of geometrical structures that generalize to phase space the spacetime metric and the $B$-field (where the $B$-field originates from the symplectic structure $\omega$). In fact, in the metastring  we have not one but {\it two} notions of a metric.
The first metric $\eta$  is a neutral metric that defines a bi-Lagrangian structure and allows to define the classical spacetime as a Lagrangian sub-manifold\footnote{We remind the reader that in symplectic geometry, a Lagrangian subspace is a half-dimensional submanifold of phase space upon which the symplectic form pulls back to zero. In simple terms, a Lagrangian submanifold might be the subspace coordinatized by the $q$'s within the phase space coordinatized by $q$'s and $p$'s.} --- more precisely, the classical spacetime is defined as a null subspace for $\eta$ which is also Lagrangian for $\omega$. 
The second metric $H$ is a metric of signature $(2,2(D-1))$ that encodes the geometry along the classical spacetime (of dimension $D$) as well as the transverse energy-momentum space geometry.
In this formulation, T-duality exchanges the Lagrangian sub-manifold with its image under $J=\eta^{-1}H$.
Classical metastring theory is defined by the following action \cite{Freidel:2015pka} 
which realizes the above comments about quantum field theory in the modular polarization for the 
special case of a two-dimensional world-sheet 
quantum field theory\footnote{See also \cite{Tseytlin:1990va}.}
\begin{equation}\label{metastringAction}
\hat{S}=  \frac{1}{4\pi}\int_{\Sigma}d^2\sigma \Big(  \pa_{\tau}{\X}^{A} (\eta_{AB}+\omega_{AB}) (\X) \pa_{\sigma}\X^{B} -  
  \pa_{\sigma}\X^{A}  H_{AB} (\X) \pa_{\sigma}\X^{B}\Big) ,
\end{equation}
where $\X^A$ are dimensionless coordinates on phase space and the fields $\eta, H,\omega$ are all dynamical (i.e., in general dependent on $\X$)  phase space fields. In the context of a flat metastring we have 
 constant $\eta_{AB}$,  $H_{AB}$
 and $\omega_{AB}$
\be\label{etaH0}
	\eta_{AB} \equiv \left( \begin{array}{cc} 0 & \delta \\ \delta^{T}& 0  \end{array} \right),\quad
H_{AB} \equiv  \left( \begin{array}{cc} h & 0 \\ 0 &  h^{-1}  \end{array} \right),
\quad \om_{AB} = \left( \begin{array}{cc} 0 & \delta \\ -\delta^{T}& 0  \end{array} \right) ,
\ee
where $\delta^{\mu}_{\nu}$  
 is the $d$-dimensional identity matrix and $h_{\mu\nu}$ is the $d$-dimensional Lorentzian metric, $T$ denoting transpose.

{In view of our general comments regarding the modular polarization (quantum spacetime polarization) in quantum field theory, the metastring sheds new light on
some old questions regarding the continuum limit of string theory \cite{Klebanov:1988ba} as well as the Wilsonian approach to string theory \cite{Hughes:1988bw}.}
In the metastring formulation \cite{Freidel:2015pka} it  is convenient, as suggested by the double field formalism \cite{Hull:2009mi}, 
to introduce  dimensionless  coordinates $\X^{A}\equiv (X^{\mu}/\lambda ,P_{\mu}/\varepsilon )^{T}$  on phase space\footnote{{
See also \cite{Minic:1991rk} as an early example of a phase space
formulation of string theory (and thus of relative, observer-dependent, locality in string theory).} },
or equivalently, $\X^{A}\equiv (x^a , \tilde{x}_a)^{T}$,
where $\lambda$ and $\epsilon$ represent the fundamental spacetime and energy-momentum scales.
Here $\hbar = \lambda \epsilon$ and $\alpha'= \frac{\lambda}{\epsilon}$.
Given a pair $(H,\eta)$ it is natural to consider the operator
$J\equiv \eta^{-1}H$. The consistency of string theory requires 
$J$ to be a chiral structure, that is, a real structure ($J^2=1$) compatible with $\eta$, implying that $J$ is an $O(D,D)$ transformation (realizing generalized T-duality in target space). These three structures, the symplectic $Sp(2D)$ $\omega$, the $O(D,D)$ $\eta$ and the $SO(2,2(D-1))$ $H$, define the
new concept of Born geometry \cite{Freidel:2013zga, Freidel:2014qna, Freidel:2015pka, Freidel:2015uug, Freidel:2016pls} (see also \cite{Freidel:2017yuv}) which unifies the complex geometry of quantum theory with
the metrical geometry of general relativity and the symplectic geometry of canonical Hamiltonian dynamics 
\cite{Gibbons:1991sa}. 
Note that in the phase space formulation the local phase space coordinates $\X$ are {\it quasiperiodic}
$
\X^A (\sigma + 2\pi) = \X^A(\sigma) + \Delta^A,
$
where $\Delta^A$ is the corresponding quasiperiod (which either vanishes for the canonical Polyakov string or
is given by the winding number in the usual treatment of T-duality on compact spaces).

The worldsheet formulation of the metastring is chiral. Thus, even though the fields are
doubled the central charges (left and right) are $c_L= c_R=D$ and we still have $D=26$ for criticality.
The metastring is not manifestly invariant under the worldsheet Lorentz transformations and it contains monodromies $\X^A (\sigma + 2\pi) = \X^A(\sigma) + \Delta^A$.
The usual Polyakov string can be obtained by integrating out the dual $\tilde{X}$,
for constant $\eta$ and $H$ backgrounds, and by supposing that the monodromies are in the kernel of $(\eta - \omega)$.
T-duality is implemented in target space by the action of the chiral $J$ operator ($J \equiv \eta^{-1} H$, $J^2=1$):
$\X \to J(\X)$.

The target space of the metastring is not spacetime, but, to first order, a chiral phase space $\cal{P}$ equipped by the 
symplectic structure $\omega$, and the bilagrangian structure, and in particular, the polarization metric $\eta$ which relates to the symplectic connection of the
Fedosov deformation quantization \cite{Fedosov:1996fu} and thus leads to the star product of deformation quantization, and finally, the quantum $H$ metric which
relates to the complex structure in the context of geometric quantization \cite{geoq}, leading to the concept of Hilbert spaces.
This classical Born geometry implements the ideas of Born duality in string theory \cite{Freidel:2013zga, Freidel:2014qna}.

The classical equations of motion of the metastring 
$
\partial_{\tau} \X^A - (J \partial_{\sigma} \X)^A=0,
$
implies the relation between momenta and monodromies $2 \pi P = J(\Delta)$.
There is {\it soldering} between worldsheet null coordinates $\sigma^{\pm} \equiv \sigma \pm \tau$ 
and the chiral target space structure
$
\partial_{\pm} \X^A - (P_{\pm} \X)^A=0,
$
where the chiral projector is defined as $2 P_{\pm} = (1\pm J)$.
This allows us to liberate the left geometry from the right geometry (which is reminiscent of twistor theory).
The careful analysis of the metastring action \cite{Freidel:2015pka} shows that its symplectic form is
$
\Omega = \frac{1}{4\pi} \int \delta \X^A \eta_{AB} \nabla_{\sigma} \delta \X^B,
$
where $\nabla$ is the generalized Fedosov connection found in the Fedosov deformation quantization approach \cite{Fedosov:1996fu}.

Also, the operator product expansion of the metastring vertex operators 
$
V_k = \epsilon_k e^{i \K\X},
$
(i.e. modular variables) lead to the restriction of $\K$ on a double Lorentzian integral lattice $\Gamma$, that by modular invariance, must be
self-dual. These exist in $D=2 mod(8)$, and are unique. Criticality gives a very unique lattice
$
\Gamma = \Pi_{1,25} \times \Pi_{1,25}.
$
This fact, in turn, leads to the large symmetry structure found by Borcherds in the
study of the monstrous moonshine \cite{Borcherds:1983sq}
\footnote{For a string theory related discussion, see \cite{Moore:1993zc}.}.

As already noted, the metastring is chiral. This requires the introduction of a
preferred worldsheet time coordinate which is fundamentally
Lorentzian \cite{Freidel:2015pka}\footnote{{See also Witten's treatment of Feynman's $i \epsilon$ prescription in string theory, which requires doubling
of $X$s and the Lorentzian world-sheet, as required by metastring theory \cite{Witten:2013pra}.}}.
How can this be consistent with modular invariance?
The answer is given by employing the Giddings-Wolpert-Krichever-Novikov construction 
\cite{Giddings:1986rf}: given a Riemann surface, provided a choice of
local coordinates around punctures is labeled by one scalar, there exists a unique Abelian differentional $e$ with imaginary periods.
The real part of this Abelian differential is the modular invariant time $\tau = Re(e)$. The zeros of $e$ represent
interaction points where the worldsheet Lorentzian cones double. Cutting the Riemann surface along the real trajectory of 
$e$ we obtain a string decomposition of the surface. 
The Nakamura graphs \cite{Nakamura} encode this decomposition and give a very effective cell decomposition of moduli space.
Thus Nakamura graphs are the natural Feynman diagrams for closed strings \cite{Freidel:2014aqa}\footnote{{
Note that the metastring is bosonic, and that supersymmetry (and the superstring) can be viewed as emergent from this more fundamental formulation (see \cite{Englert:1986na}).
In some sense, fermions can be viewed as defects in modular spacetime. The bosonic degrees of freedom associated with quanta of interactions can
be viewed as deformations of modular spacetime cells.}}.

\subsection{Non-commutativity in quantum gravity/metastring theory}

The metastring formulation points to an unexpected fundamental non-commutativity of closed string theory, that we address in what follows.
It is well established \cite{Polchinski:1998rq} that the structure of the zero mode algebra of the compactified closed string 
depends on a lattice of momenta $(\Lambda,2\eta)$ which is integral and self-dual
with respect to a neutral metric: a so-called Narain lattice \cite{Narain:1985jj}. 
In 
\cite{Freidel:2013zga, Freidel:2014qna, Freidel:2015pka, Freidel:2015uug, Freidel:2016pls, Freidel:2017xsi, Freidel:2017wst, Freidel:2017nhg}
we have refined this structure and we have shown that in fact the kinematical structure of the string zero modes depends on a {\it para-hermitian lattice}:  a triple $(\Lambda, \eta, \omega)$, where
$\Lambda$ is a subgroup of $\mathbb{R}^{2d}$ that describes the lattice of wave-covectors $\lambda \K$, with $\lambda$ the string length, $\eta$ is a neutral metric, a symmetric bilinear form of signature $(d,d)$, and  $\omega$ is an invertible two-form. This structure needs to satisfy two compatibility conditions: 
first, the lattice  $\Lambda$ is assumed to be integral with respect to the para-hermitian structure, i.e., 
$(\eta\pm \omega)(\lambda \K,\lambda \K') \in \mathbb{Z},$
 for $\lambda \K,\lambda \K'\in \Lambda$. 
 Second, the metric $\eta$ and the 2-form $\omega$ must be compatible, in the sense that $\eta^{-1}\omega:=K$ is a product structure, that satisfies the condition $K^2=1$. 

 These two conditions are a consequence of mutual locality on the worldsheet (i.e. worldsheet causality).
It is clear that if $(\Lambda,\eta,\omega)$ is a para-hermitian lattice, then $(\Lambda,2\eta)$ is a Narain lattice, so the kinematical structure that we  highlight  is a refinement of the usual one.
The extra information is contained in the 2-form $\omega$.
This form does not enter expressions for the spectrum or the partition function and this why it is usually ignored.
It does enter however crucially in the definition of the vertex operator algebra and parameterizes what is usually referred to as a cocycle.
The role of $\omega$ is to promote the zero mode double space ${\cal P} \simeq \R^{2d}$ dual to $\mathbb{R}[\Lambda]$ to the status of phase space: ${\cal P}$ should be viewed as a  symplectic manifold.
At the quantum level,  both geometrical structures $\eta$ and $\omega$ enter  in the commutation relations of  string operators. $\omega$ controls the non-commutativity of the zero-modes while $\eta$ controls the non-commutativity of the string oscillator modes. This can be seen if one introduces a double notation for the string coordinate $\mathbb{X}(\s)$ that includes the string map $X$ and its dual $\tilde{X}$. The string commutation relations, were derived in \cite{Freidel:2017wst, Freidel:2017nhg}
\be
[\X^A(\s),\X^B(\s')]=  2i\lambda^2 \left[ \pi \omega^{AB}-\eta^{AB}\theta(\sigma-\s')\right],
\ee
where $\theta(\s)$ is the staircase distribution, i.e., a solution of $\theta'(\s)=2\pi\delta(\s)$; it is odd and  quasi-periodic with period   $2\pi$.

Following standard practice, all indices are raised and lowered using  $\eta$ and $\eta^{-1}$. 
The momentum density operator is given by ${\mathbb{P}}_A(\s)= \frac{1}{2\pi\alpha'} \eta_{AB}\pa_\s \X^B(\s)$ and the previous commutation relation implies that it is conjugate to $\X^A(\s)$. The two-form 
$\omega$ appears when one integrates this canonical commutation relation to include the zero-modes, the integration constant being uniquely determined by worldsheet causality. Denoting by $(\hat\X,\hat{\mathbb{P}})$ 
the zero mode components of the string operators $\X(\s)$ and ${\mathbb{P}}(\s)$  we simply have that 
\be
[\hat{{\mathbb{P}}}_A,\hat{{\mathbb{P}}}_B]=0,%\qquad 
[\hat{\X}^A,\hat{{\mathbb{P}}}_B]=i\hbar \delta^A{}_B,%\qquad 
[\hat{\X}^A,\hat{\X}^B ]=  2\pi i\lambda^2 \omega^{AB}.
\ee
This is a deformation of the doubled Heisenberg algebra involving the string length $\lambda$ as a deformation parameter.

So far we have assumed that the background is trivial, with the fields $(\eta,\omega)$  {\it constant} and given by 
%\beq\label{flatetaom}
$\eta(\K,\K') =k \cdot \tilde{k}' + \tilde{k}\cdot k',$ and %\qquad
$\omega(\K,\K')=k \cdot \tilde{k}' - \tilde{k}\cdot k'.$
%\eeq
As shown in \cite{Freidel:2017wst}, we can turn on non-trivial backgrounds encoded into $\omega$ by changing the O$(d,d)$ frame $\X \to O\X$. This change of frame preserves $\eta$ but  transforms $\omega$. Any constant $\omega$ can be obtained this way. Since $\omega$ has an interpretation as the symplectic form on the space of $\X$'s, modifying $\omega$ affects the commutation relations\footnote{The algebraic structure that we are working with here has an analogy in electromagnetism in the presence of monopoles. In that analogy, the string length becomes the magnetic length, and the form $\omega$ becomes the magnetic field.  Another analogy occurs in quantum Hall liquids, the algebra being the magnetic algebra of the lowest Landau level.
}
$
[\hat{\X}^A,\hat{\X}^B]= 2\pi i\lambda^2\Pi^{AB},$, with $ \Pi^{AB} \omega_{BC}= \delta^A{}_C, $
where we have introduced the Poisson tensor $\Pi=\omega^{-1}$. 

 For instance, under a constant $B$-field transformation 
 $\X=(x^a,\tx_a) \mapsto (x^a, \tx_a + B_{ab} x^b)$, 
% $\K=(\tk^a,k_a) \mapsto (\tk^a, k_a + B_{ab} \tk^b)$, 
 the trivial symplectic form $\omega(\K,\K')=k \cdot \tilde{k}' - \tilde{k}\cdot k'$ is mapped onto
$
\omega(\K,\K')=  k_a  \tk'^a - k'_a \tk^a - 2 B_{ab} \tk^a \tk'^b,
$
and the commutators read
 \beq\label{Bcommrel}
[\hat{x}^a,\hat{x}^b]=0,\qquad 
[\hat{x}^a,\hat{\tx}_b]=2\pi i\lambda^2 \delta^a{}_b,\qquad 
[\hat{\tx}_a,\hat{\tx}_b]=-4\pi i\lambda^2 B_{ab}.
\eeq
We see that the  effect of the $B$-field is to render the dual coordinates non-commutative (and that the $B$-field originates from the symplectic structure $\omega$). 
More generally, we can parameterize an arbitrary $O(d,d)$ transformation as $g=e^{\hat{B}}\hat{A}e^{\hat\beta}$, where $\hat {A}\in GL(d)$ and $e^{\hat{B}}=\tiny\begin{pmatrix}{\bf 1}&0\cr B & {\bf 1}\end{pmatrix}$ and $e^{\hat{\beta}}=\tiny\begin{pmatrix}{\bf 1}&\beta\cr {\bf 0} & {\bf 1}\end{pmatrix}$ are nilpotent. $e^{\hat B}$ is the $B$-field transformation discussed above, and is associated with the usual $B$-field deformation in string theory. We note that the transformation of $(x^a,\tx_a)$ given above does not modify $x^a$, and thus fields that depend only on $x^a$ are unmodified. The $\beta$-transformation on the other hand corresponds to the map $(x^a,\tx_a)\mapsto (x^a+\beta^{ab}\tx_b,\tx_a)$. Equivalently, it has the effect of mapping the symplectic structure to
$
\omega(\K,\K')=  k_a  \tk'^a - k'_a \tk^a + 2 \beta^{ab} k_a k'_b,
$
and yields commutation relations
 \beq\label{betacommrel}
[\hat{x}^a,\hat{x}^b]=4\pi i\lambda^2 \beta^{ab},\qquad 
[\hat{x}^a,\hat{\tx}_b]=2\pi i\lambda^2 \delta^a{}_b,\qquad 
[\hat{\tx}_a,\hat{\tx}_b]=0.
\eeq
Dramatically, the coordinates that are usually thought of as the spacetime coordinates have become themselves non-commutative. 
Since this is the result of an $O(d,d)$ transformation, we know that it can be thought of in similar terms as the $B$-field; these are related by T-duality. We are familiar with the $B$-field background because we have, in the non-compact case, a fixed notion of locality in the target space theory. However, in the non-geometric $\beta$-field background, we do not have such a notion of locality but we can access it through T-duality\footnote{Note that the dilaton can be understood as coming from the volume of phase space (see  \cite{Boffo:2019zus}).
In general, for varying $B$-backgrounds we encounter non-associativity as well \cite{nonassoc}, and the proper closure of such 
non-commutative and non-associative structure is ensured by the equations of motion for the data of Born geometry, that is, the generalized Einstein equations.}.

We note that this intrinsic non-commutativity of string theory can be also explicitly
illustrated via a simple closed string product, equivalent to the splitting-joining
interaction of the pants diagram, that respects this non-commutativity and is covariant
with respect to T-duality \cite{Freidel:2017nhg}. This offers new insights on the relationship between
closed and open strings, and the non-perturbative formulation of closed string theory
in terms of open strings and even more fundamental (and non-commutative)
partonic degrees of freedom. Given the mechanism of tachyon condensation in the
open string sector \cite{Sen:1998sm} and this fundamental relation between open and closed strings,
we expect that a similar solution of the ``tachyon problem" should exist in the closed
string sector as well.

\subsection{Non-perturbative formulation of quantum gravity}

Let us recapitulate: How does the metastring approach \cite{Freidel:2013zga, Freidel:2014qna, Freidel:2015pka, Freidel:2015uug, Freidel:2016pls, Freidel:2017xsi, Freidel:2017wst, Freidel:2017nhg, Freidel:2018apz, Freidel:2019jor} compare to the usual view of string theory?
From the classic textbook treatment of string theory \cite{Polchinski:1998rq}
we know that there exists a fundamental relation between world-sheet conformal
field theory (CFT) and target spacetime geometry, and, in particular, that
the beta function for the background spacetime metric is the Ricci tensor to leading
order, and so, world-sheet conformal invariance implies
the vacuum Einstein equations.
But string theory has other background field and fluxes and moduli.
Can we write the general CFT as a generalized Ricci flow? 
This is precisely the achievement of Double Field Theory (DFT) \cite{Hull:2009mi}.

In the metastring formulation the target space is not the index space (the labels of the background effective field).
Effective fields are associated with the level matching constraints or mutual locality (world sheet causality).
In DFT that requirement appears as the section condition.
Instead of this section condition, in the metastring we have a more general background spacetime (modular spacetime) for the string,
because of the intrinsic non-commutativity, stemming from the general quasi-periodicity of the worldsheet $X$  fields.
Thus even though the currents $dX$ are periodic, the fields $X$ are not, and this leads to edge modes in
the evaluation of the string symplectic form, which ultimately leads to intrinsic non-commutativity of the metastring.
Similarly, mutual locality of string vertex operators implies that, in general, they furnish a representation of 
a Weyl-Heisenberg algebra. Note that the metastring treats the Hamiltonian and diffeomorphism constraints together,
on the same footing,
and thus instead of solving the differomorphism constraint (via level matching and the strong constraint) we
arrive at a completely new background interpretation: the modular spacetime.

The new ingredient of the metastring, as compared to DFT is the symplectic structure that controls intrinsic non-commutativity.
This in turn, with the $O(d,d)$ structure (also associated with level matching and the string diffeomorphism constraint)
and the conformal (double metric) structure (associated with the string Hamiltonian constraint)
defines the Born geometry of the metastring.
Note that now we have a new spacetime that defines the habitat of string theory, associated with the maximally
commuting subalgebras of the Heisenberg algebra of the intrinsic string non-commutativity.
This quantum Lagrangian is modular spacetime of the metastring.
This in turn directly relates to the quantum spacetime formulation (modular polarization) of quantum theory
in general. The 1d modular line is the 2d torus that is compact and not simply connected.
Note the Lorentz covariance of modular spacetime, because the intersection of the symplectic, $O(d,d)$ and
the double metric structures leads to the Lorentz group.

In the context of general curved Born geometry the symplectic structure is not closed, and we have non-associativity \cite{nonassoc}.
Finally, supersymmetry (SUSY) is in principle emergent (as in the constructions of the superstring from the
bosonic string \cite{Englert:1986na}). The uniqueness of the connection in generalized geometry can be fixed by the phase space
structure, instead of SUSY.

In thinking about this new framework for quantum gravity, the teleparallel formulation \cite{einstein} of
gravity appears as the natural language, because of the inherent ``flatness'' of T-duality (which might be
important in the cosmological context for the geometry of classical spacetime at large scales). In this formulation,
gravity is described as a Yang-Mills theory, and it is put on the same footing as the matter sector.
(It is an interesting question to understand whether teleparallel equations follow from the requirement of the associativity of the symplectic form?
That Einstein equations, as well as other equations for the massless modes of the string, follow from
the closure of a symplectic-like form is the hallmark of the Bowick-Rajeev geometric quantization approach to string theory
\cite{Bowick:1986rc}.)

The metastring offers a new view on the fundamental question of a non-perturbative formulation of quantum gravity \cite{Freidel:2013zga, Freidel:2014qna, Freidel:2015pka, Freidel:2015uug, Freidel:2016pls, Freidel:2017xsi, Freidel:2017wst, Freidel:2017nhg, Freidel:2018apz, Freidel:2019jor}.
Note that the world-sheet can be made modular in our formulation, with the doubling of $\tau$ and $\sigma$, so
that  $\X (\tau, \sigma)$ can  be
in general viewed as an infinite dimensional matrix (the matrix indices coming from the Fourier components of the
doubles of $\tau$ and $\sigma$). Then the corresponding metastring action should look like
\be
\int Tr [ \partial_{\tau} \X^A \partial_{\sigma} \X^B (\omega_{AB} + \eta_{AB}) - 
\partial_{\sigma} \X^A H_{AB} \partial_{\sigma} \X^B ] d \tau d \sigma ,
\ee
where the trace is over the matrix indices.
Then we could associate the natural partonic degrees of freedom with matrix entries.
We arrive at a {\it non-perturbative quantum gravity} by replacing the sigma derivative with a
commutator involving one extra $\X^{26}$ (with $A=0,1,2,...,25$)\footnote{That the canonical world-sheet of string theory
might become non-commutative in a deeper, non-perturbative formulation, was suggested in \cite{Atick:1988si}.}
\be
\partial_{\sigma} \X^A \to [\X^{26}, \X^A] .
\ee
This dictionary suggests the following fully interactive and non-perturbative formulation of metastring theory
in terms of a (M-theory-like) matrix model form of the above metastring action (with $a,b,c=0,1,2,...,25, 26$ )
\be
\int Tr ( \partial_{\tau} \X^a [\X^b, \X^c] \eta_{abc}  - 
H_{ac} [\X^a, \X^b] [\X^c, \X^d] H_{bd}) d \tau ,
\ee
where the first term is of a Chern-Simons form and the second of the Yang-Mills form, and $\eta_{abc}$ contains both 
$\omega_{AB}$ and $\eta_{AB}$.
{\it This is then the non-perturbative gravitization of the quantum.}\footnote{In the case of the non-perturbative matrix theory like formulation of the metastring (and quantum gravity)
the matrices emerge from the modular world-sheet, and the fundamental commutator from the
Poisson bracket with respect to the dual world sheet coordinates (of the modular/quantum world sheet) - that is,
quantum gravity ``quantizes'' itself, and thus quantum mechanics originates in quantum gravity (see also, \cite{Adler:2002fu}).
This formulation should be distinguished from Penrose's ``gravitization of the quantum'' and gravity induced ``collapse of the wave function'' \cite{Penrose:2014nha}.
Also note some similarity of the metastring formulation, in its intrinsic non-commutative form, to
the most recent proposal by Penrose regarding ``palatial'' twistor theory 
\cite{Penrose:2015lla}.) }

When discussing a non-commutative phase of string theory it is natural to invoke the IIB matrix model~\cite{Ishibashi:1996xs}, which describes $N$ D-instantons (and is by T-duality related to the Matrix model of M-theory~\cite{Banks:1996vh}).
Given our new viewpoint we can suggest a {\it new covariant} non-commutative matrix model formulation of string theory, as a theory of quantum gravity, by writing in the large $N$ limit $\pa_\s\X^C = [\X, \X^C]$ 
(and similarly for $\pa_{\tau}\X^B$) in terms of commutators of two (one for $\pa_\s\X^C$ and one for $\pa_{\tau}\X^C$) extra $N\,{\times}\,N$ matrix valued chiral $\X$'s.
{\it Notice that, in general, we do not need an overall trace, and so the action can be viewed as a matrix, rendering the entire non-perturbative formulation as
purely quantum in the sense of the original matrix formulation of quantum mechanics (Born-Jordan and Born-Heisenberg-Jordan \cite{history}) }:
\begin{equation}
 \S_{\text{ncF}}\,{=}\,\frac{1}{4\pi} %\mathrm{Tr} 
[{\X}^{a},  {\X}^{b}] [{\X}^{c},  {\X}^{d}]  f_{abcd},
% \label{e:}
\end{equation}
where instead of $26$ bosonic $\X$ matrices one would have $28$, with supersymmetry emerging in 10($+$2) dimensions from this underlying
 bosonic formulation. (This would be a non-commutative matrix model formulation of F-theory.)
By T-duality, the new covariant M-theory matrix model reads as 
\begin{equation}
 \S_{\text{ncM}}\,{=}\,\frac{1}{4\pi}  
\int_{\tau} %\mathrm{Tr} 
\big(\pa_{\tau} \X^i [{\X}^{j},  {\X}^{k}] g_{ijk} - [{\X}^{i},  {\X}^{j}][{\X}^{k},  {\X}^{l}] h_{ijkl}\big),
% \label{e:}
\end{equation}
with 27 bosonic $\X$ matrices, with supersymmetry emerging in 11 dimensions\footnote{In this approach
holography \cite{tHooft:1993dmi} (such as AdS/CFT \cite{Maldacena:1997re}, which can be viewed as a ``quantum Jarzynski equality on the space of geometrized RG flows'' \cite{Minic:2010pw}) is emergent in a particular ``extensification''
of quantum spacetime.}. 
The relevant information about
 $\w_{AB},\eta_{AB}$ and $H_{AB}$ is now contained in the new dynamical backgrounds $f_{abcd}$ in F-theory, and $g_{ijk}$ and $h_{ijkl} $ in M-theory\footnote{This offers a new formulation of covariant Matrix theory in the M-theory limit\cite{Minic:1999js},
which is essentially a partonic formulation - strings emerge from partonic constituents in a certain limit. This new matrix formulation is fundamentally bosonic and thus it is reminiscent of bosonic M-theory \cite{Horowitz:2000gn}.
The relevant backgrounds $g_{ijk}$ and $h_{ijkl} $ should be determined by the matrix RG equations.
Also, there are lessons here for the new concept of ``gravitization of quantum theory'' as well as the idea that dynamical Hilbert spaces or 2-Hilbert spaces (here represented by matrices) are fundamentally needed
in quantum gravity \cite{twohilbert}.}.

This matrix like formulation should be understood as a general non-perturbative formulation of 
string theory. In this partonic formulation closed strings are collective excitations, in turn constructed from the product of open string fields. The observed classical spacetime emerges as an ``extensification'' \cite{Freidel:2013zga, Freidel:2014qna, Freidel:2015pka, Freidel:2015uug, Freidel:2016pls, Freidel:2017xsi, Freidel:2017wst, Freidel:2017nhg, Freidel:2018apz, Freidel:2019jor}, in a particular limit, out of the basic building blocks of quantum spacetime.
Their remnants can be found in the low energy bi-local quantum fields, with bi-local (metaparticle) quanta, to which we now turn.

\subsection{Metastrings, quantum fields and metaparticles}

What is the effective description of closed strings that incorporates the above intrinsic non-commutativity? For a closed sting on a circle of radius $R$ (where the dual radius $\tilde{R}$, is defined as $R \tilde{R} = 2 \lambda^2$ and the respective winding integers are $n$ and $w$) this effective description is captured by the generalized field \cite{Freidel:2017wst, Freidel:2017nhg}
\beq\label{Gfield}
\Phi(x,\tx)\equiv \sum_w \Phi_w(x) e^{iw\tx/\tR}.
\eeq 
This meshes well with the observation \cite{Freidel:2017wst, Freidel:2017nhg} that the string product is essentially a representation of the Heisenberg group, which suggests that one should  consider the ``quantization'' map 
\beq\label{Gfieldw}
\Phi(x,\tx) \to \hat{\Phi} =\sum_w \Phi_w(\hat{x}) e^{iw\hat{\tx}/\tR},
\eeq from generalized fields to  non-commutative fields.\footnote{Here, we have chosen a specific operator ordering. Given this ordering, the mapping is well-defined and consistent with the string product.} 
Under this map the T-duality transformation becomes ``localized'' and is expressed as the exchange of $\hat{x}$ with $\hat{\tilde{x}}$. The T-dual expression is given by \cite{Freidel:2017wst, Freidel:2017nhg}
\beq\label{Tfield}
\hat\Phi=\sum_n e^{in\hat{x}/R} \Phi_n(\hat{\tx}-\pi n\tR)
=
\sum_n   \Phi_n(\hat\tx) e^{in\hat{x}/R},
\eeq 
which has a similar form to (\ref{Gfieldw}).
We see that the non-commutativity of $\hat{x}$ with $\hat\tx$ allows one to reabsorb all the shifts in terms of a simple reordering that exchanges  $\hat{x}$ with $\hat\tx$ and is the expression of T-duality.  The ``quantized'' field is simply expanded in terms of modes as 
\beq
\hat\Phi \equiv \sum_{w,n}  e^{in\hat{x}/R}  \Phi(n,w) e^{iw\hat{\tx}/\tR}. 
\eeq

It is useful at this point to generalize the construction to higher dimensional tori. This can be done in a straightforward manner by introducing the modes
$\K^A=(\tk^a, k_a )$, generalizing  $ ( w/\tilde{R},n/R)$. 
The integrality condition for the lattice $\Lambda$ of admissible modes $\K,\K'\in\Lambda$ reads in this notation as\footnote{In the one dimensional case where $\K=(w/\tilde{R},n/R)$ this follows directly from 
$(\eta+ \omega)(\lambda \K,\lambda \K') = n w'$ and similarly $(\eta- \omega)(\lambda \K,\lambda \K')= wn' $, given that $n,n',w,w'\in\mathbb{Z}$.}
$
(\eta \pm \omega)(\lambda \K,\lambda \K')\in \mathbb{Z}.
$
We now write $\Phi(\K)=\langle \K|\Phi\rangle$ with the  ordering chosen as
$
\langle \K| = \langle 0| \hat{U}_{-\K}$, where $ \hat{U}_{\K} \equiv e^{i k\cdot \hat{x}} e^{i\tilde{k} \cdot \hat{\tx}}.
$
This ordering can be seen to be related to the choice of an O$(d,d)$ frame, where we place the operator associated with ${x}$ on the left and the operator associated with the dual space $\tilde{x}$ on the right. 
The key point is that this choice of frame is entirely encoded into the choice of symplectic potential $\omega$ and the vertex operator can be covariantly written in terms of $\K=(\tilde{k},k) $ and $\X=(x,\tilde{x})$ as 
\be 
\hat{U}_{\K} = e^{\frac{i}{2} (\eta + \omega)(\K,\hat\X)}e^{\frac{i}{2} (\eta - \omega)(\K,\hat\X)}.
\ee
Given this notation we can write the string product covariantly as \cite{Freidel:2017wst, Freidel:2017nhg}
\beq \label{proddual}
(\Phi\circ\Psi)(\K)
&=&\sum_{\K'+\K'' =\K }  \Phi(\K') e^{ i\pi(\eta- \omega)(\lambda\K',\lambda\K'')}\Psi(\K'').
\eeq
The non-commutativity of the string product is encoded in terms of a $\pi$-flux due to $\omega$. As it turns out the phase factor is exactly the same as the cocycle factor
$\epsilon(\K,\K')= e^{ i\pi (\eta- \omega)(\lambda\K,\lambda\K')}$ that appears in the 
definition of the vertex operator product \cite{Freidel:2017wst, Freidel:2017nhg}.

Thus, quantum gravity, in the guise of metastring theory, produces at low energy bi-local quantum fields, with intrinsic non-commutatitivity
$\phi(x, \tx)$ where $[x, \tx] = i \lambda^2$. As we have already discussed in section 3, and as we will see more explicitly in the following section 5,
these bi-local fields are doubled, and their proper formulation requires a double scale renormalization group (RG) found in the
context of non-commutative field theory \cite{Douglas:2001ba} 
(which in certain models lead to a finite non-perturbative renormalization).
Such doubled bi-local fields ($\phi (x, \tx)$ and $\tilde{\phi} (x, \tx)$) have metaparticle excitations to be discussed in the next section.
One can view such (meta)fields as low energy manifestations of the metastring field.
These intrinsically non-commutative quantum fields can also be understood to arise from the
representation of the symmetry groups associated with Born geometry.
They should be relevant both in the high energy context (see the discussion that follows on dark matter and 
dark energy\footnote{Even the Standard Model of particle physics (coupled to Einstein's gravity) exhibits hidden non-commutative geometry (NCG), as discussed in \cite{connes} (for  
recent reviews and references, consult \cite{Chamseddine:2019fjq}), 
and this implies some unique phenomenological consequences
\cite{Aydemir:2013zua}. The NCG action should be compared to (34), by replacing the commutators with
the NCG Dirac operator.}) 
as well as in condensed matter physics (as new quantum order parameters for 
highly entangled and strongly correlated phases of quantum matter).

To summarize, the above manifestly T-duality covariant formulation of closed strings (i.e. the metastring) implies intrinsic non-commutativity of zero-modes. It is thus instructive to formulate a particle-like limit of the metastring 
that we call the {\it metaparticle} \cite{Freidel:2018apz}. 
Given the form for the symplectic structure of the zero modes derived in section 4. of \cite{Freidel:2017wst} (equation (67) of that paper, without the contribution coming from string oscillators), the action $S \equiv \int d\tau L$ of the metaparticle is governed by the following Lagrangian  
(implied by the symplectic structure of the closed string) \cite{Freidel:2017wst, Freidel:2017nhg}
\beq
 L = p_\mu\, \dot x^\mu +\tilde p^\mu\, \dot{\tilde x}_\mu +\alpha' p_\mu\, \dot{\tilde p}^\mu - \frac{N}{2}\left(p_\mu p^\mu +\tilde p_\mu \tilde p^\mu - m^2\right) +{\tilde{N}}\left(p_\mu \tilde p^\mu - \mu \right),
\eeq
where $N$ and $\tilde{N}$ are the Lagrange multipliers for the two constraints that follow from the Hamiltonian ($H\equiv \pa_{\sigma}\X^{A} H_{AB} \pa_{\sigma}\X^{B} =0$) and diffeomorphism constraints ($D\equiv \pa_{\sigma}\X^{A}\eta_{AB} \pa_{\sigma}\X^{B} =0$) of the metastring \cite{Freidel:2015pka, Freidel:2017xsi}.

Note that the usual particle limit is obtained, at least classically, by taking $\mu \to 0$ and $\tilde p \to 0$.
The theory of metaparticles can be viewed as the theory of the zero modes of the closed string, which fully takes into account
its intrinsic non-commutativity. Given the form of the above Lagrangian, the metaparticle looks like two particles that are entangled through a Berry
phase-like $p_\mu\, \dot{\tilde p}^\mu$ factor. The metaparticle is fundamentally non-local, and thus it should not be associated with effective local field theory.
In particular, by looking at the metaparticle constraints
$
p^2 + {\tilde{p}}^2 = m^2$ and $ p \tilde{p} = \mu,
$
we note that the momenta $p$ and $\tilde{p}$ can be, in principle, widely separated.
For example, if $m$ is of the order of the Planck energy, and $\mu$ of the order of one $TeV$ (which could be understood
as a characteristic particle physics scale), then the momentum $p$ can be of the order of the Planck energy, and the momentum $\tilde{p}$ of the vacuum energy scale. Thus the metaparticle theory is able to naturally relate widely separated scales, which transcends the usual reasoning based on Wilsonian effective field theory (and should be relevant for the naturalness and hierarchy problems).

\section{Quantum gravity, metaparticles and dark matter}

The theory of metaparticles (the low energy remnants of the metastring, and as such, the low energy remnants of quantum gravity) can thus be defined by the following world-line action \cite{Freidel:2018apz} 
\begin{equation}\label{1}
S \equiv \int_0^1 d\tau [p \dot x +\tilde p \dot{\tilde x} + \alpha'\, p \dot{\tilde p} - \frac{N}2\left(p^2 +{\tilde p}^2 + m^2\right) +{\tilde N}\left(p \tilde p - \mu \right)]\,.
\end{equation}
Here the signature $(+,-,\ldots,-)$ and the contraction of indices are implicitly assumed.
At the classical level, theory of metaparticles is a world-line theory with the usual reparameterization invariance and two additional features \cite{Freidel:2018apz}.  
The first new feature of the model is the presence of an additional local symmetry, which from the string point of view corresponds to the completion of worldsheet diffeomorphism invariance. From the particle world-line point of view, 
this symmetry is associated with an additional local constraint. The second new feature is the presence of a non-trivial symplectic form on the metaparticle phase space, also motivated by string theory \cite{Freidel:2017wst, Freidel:2017nhg}. 
Because of its interpretation as a particle model on Born geometry, associated with the modular representation of quantum theory,
the space-time on which the metaparticle propagates is ambiguous, with different choices related by what in string theory we would call T-duality. 
The attractive feature of this model include world-line causality and unitarity, as well as an explicit mixing of widely separated energy-momentum scales.
The metaparticle propagator follows from the world-line path-integral defined by the above action 
and it has the following form in momentum space  \cite{Freidel:2018apz}
\begin{equation}\label{doubletramp}
G(p,\tilde p; p_i,\tilde p_i)
\sim
\delta^{(d)}(p-p_{i})\delta^{(d)}(\tilde p-\tilde p_{i})
\frac{\delta(p\cdot\tilde p-\mu)}{p^2+\tilde p^2+m^2-i\varepsilon}.
\end{equation}
The canonical particle propagator is a highly singular $\tilde p \to 0$ (and $\mu \to 0$) limit of this expression.
This propagator also predicts the following dispersion relation (in a particular gauge \cite{Freidel:2018apz}) that
can be tested in various experiments and with various probes
\be
E_p^2 + \frac{\mu^2}{E_p^2} = {{\vec{p}}}^2 + m^2.
\ee
This formulation is fully compatible with Lorentz covariance, and is a direct consequence of the consistency of
quantum theory and a minimal length (and thus Born geometry).
In general, for each particle at energy $E$ there exists a dual particle at energy $\frac{\mu}{E}$.
(This is complete analogy for the well-known prediction of antiparticles in the union of special relativity 
and quantum theory.)\footnote{Regarding the phenomenology of quantum gravity (including the phenomenology of the minimal length) another generic feature presents itself in the context of the quantum version
of the gravitational memory effect \cite{Strominger:2017zoo}, which should involve ``modular supermomentum''.
For some other tests of quantum gravity, and especially intrinsic non-locality and non-commutativity of
quantum spacetime, see \cite{mike}.}

We can also discuss the background fields that couple to the
metaparticle quanta.
Following the well-known procedure of introducing the background fields in the case of particles, by
shifting the canonical momentum by a gauge field, we might try to extend the gauging procedure to the metaparticle counterpart.
There is a possible ambiguity in this gauging which depends on which 
configuration variables one decides to work with.
If one takes $(x,\tx)$ as configuration variables,
one obtains a gauging  which could also be motivated
by the presence of a "stringy gauge field" in metastring theory \cite{Freidel:2015pka}
\beq
S\to \int \Big((p_\mu+A_\mu(x,\tilde x))\dot x^\mu+(\tilde p^\mu+\tilde A^\mu(x,\tilde x))\dot{\tilde x}_\mu+2\pi\alpha' p_\mu\dot{\tilde p}^\mu-e{\cal H}(p,\tilde p)-\tilde e{\cal D}(p,\tilde p)
\Big) .
\eeq
Indeed, if we introduce canonical momenta 
\beq
P_\mu=p_\mu+A_\mu(x,\tilde x),\qquad \tilde P^\mu=\tilde p^\mu+\tilde A^\mu(x,\tilde x) ,
\eeq
we obtain then
\bea
S&\to& \int \Big(P_\mu\dot x^\mu+\tilde P^\mu\dot{\tilde x}_\mu+2\pi\alpha' (P_\mu-A_\mu(x,\tilde x))(\dot{\tilde P}^\mu-\frac{d}{dt}\tilde A^\mu(x,\tilde x))
\\&&-e{\cal H}(P-A(x,\tilde x),\tilde P-\tilde A(x,\tilde x))-\tilde e{\cal D}(P-A(x,\tilde x),\tilde P-\tilde{A} (x,\tilde x))
\Big) ,
\eea
but, now, because of  the $\alpha'$ term we see that $\dot {\tilde A}$ contains $\dot{\tilde x}$.
The background fields $A_\mu(x,\tilde x)$ and $\tilde A^\mu(x,\tilde x)$ are the natural modular fields in this case.
(Note that this procedure explicitly realizes the general comments made in section 3 regarding quantum field theory in the modular polarization.)
Thus, the generic prediction here is the existence of a dual field $\tilde{A}$, which is entangled, with the original $A$ field\footnote{Note that the metaparticle propagator leads to 
a ``Friedel-like'' bi-local static potential \cite{jerzy} as well as non-local generalizations of quantum statistics.
The traditional statement concerning quantum gravitational effects is that they are tied to the Planck scale. However, quantum gravity can be revealed at macroscopic scales via quantum statistics. 
In particular, it was argued in \cite{Strominger:1993si} that black hole statistics is infinite statistics \cite{Greenberg:1989ty} (which is consistent with non-locality and Lorentz symmetry). 
Also, in \cite{Horava:2000tb} a statistical argument was used to argue for probable values of the cosmological constant.
More recently, such statistical arguments were used in \cite{Bianchi:2018ula} to 
analyze the black hole spin in gravitational wave observations.
See \cite{Jejjala:2007hh} for the relevance of
infinite statistics for the fine structure of dark energy. As non-local objects, metaparticles can exhibit such
quantum statistics.
}.

We expect that the correct field theoretic description of the metaparticle is in terms of the above general non-commutative (modular) field theory $\Phi(x,\tx)$ limit of the metastring \cite{Freidel:2013zga, Freidel:2014qna, Freidel:2015pka, Freidel:2015uug, Freidel:2016pls, Freidel:2017xsi, Freidel:2017wst, Freidel:2017nhg, Freidel:2018apz, Freidel:2019jor}. 
Such effective non-commutative field theory is similar in spirit to 
\cite{Douglas:2001ba}. 
Also, we note that the concept of metaparticles might be argued from the compatibility of the quantum spacetime
that underlies the generic representations of quantum theory, as discussed in \cite{Freidel:2016pls}, and thus the metaparticle might be as ubiquitous as the concept of antiparticles which is demanded by the compatibility of 
relativity and quantum theory.
The metaparticles also provide a natural route to the problem of dark matter.
To lowest (zeroth) order of the expansion in the non-commutative parameter $\lambda$ the effective action for Standard Model matter Lagrangian ($L_m$) and their duals (that could be interpreted as dark matter Lagrangian $L_{dm}$)
$S_{eff}$ takes the following form (where we have included the gravitational background as non-dynamical):
\be
S_{eff} = - \iint\! \sqrt{{g(x)}{\tilde{g}(\tx)}} [ L_m (A(x, \tilde{x})) + \tilde{L}_{dm}   ( \tilde{A}(x, \tilde{x}))+... ] .
\label{e:TsSd1}
\ee
Note that after integrating over the ``hidden variable parameters $\tilde{x}$ we get an effective theory of visible and dual (dark) matter in the observed spacetime $x$
\be
S_{eff} = - \iint\! \sqrt{{g(x)}} [ L_m (A(x)) + \tilde{L}_{dm}   ( \tilde{A}(x))+... ] .
\label{e:TsSd2}
\ee
Thus, the metaparticle can be understood as a generic message of string theory/quantum gravity for low energy physics.
Like their visible particle cousins, dark matter quanta should be detectable through their particular metaparticle entanglement to visible matter,
as indicated by equation (47): $\alpha' A_\mu \frac{d}{dt}\tilde A^\mu$ (say, for a photon and its dual).
This is a Berry-phase like term that comes from a fully covariant description, and is uniquely different
from the usual effective field theory interaction terms between visible and dark matter particles.

Such dark matter quanta are correlated to visible matter and have been discussed in the literature as
Modified dark matter \cite{Ho:2010ca}.
Modified dark matter (MDM) is, at the moment, a phenomenological model of dark matter, inspired by gravitational thermodynamics. For an accelerating Universe with positive cosmological constant $\Lambda$, certain phenomenological considerations lead to the emergence of a critical acceleration parameter related to $\Lambda$ (essentially that ``fundamental acceleration'' is just the value of $\Lambda$ expressed as acceleration $\sim cH$, where $H$ is the Hubble parameter, and thus, it is 
of the order of $10^{-10}m/s^2$). Such a critical acceleration is an effective phenomenological manifestation of MDM, and it is found in correlations between dark matter and baryonic matter in galaxy rotation curves. The resulting MDM mass profiles, which are sensitive to $\Lambda$, are consistent with observational data at both the galactic and cluster scales. In particular, the same critical acceleration appears both in the galactic and cluster data fits based on MDM \cite{Ho:2010ca}. 
Furthermore, using some robust qualitative arguments, MDM appears to work well on cosmological scales.
If the quanta of modified dark matter are metaparticles,
this may  explain why, so far, dark matter detection experiments have failed to detect dark matter particles. In particular, the natural model for MDM quanta could be
provided by the metaparticle realizations of the Standard Model particles, associated with bi-local extensions
of all Standard Model fields. Thus the baryon matter described by the Standard Model fields (the $A$ backgrounds in the
above discussion), would have natural
cousins (the $\tilde{A}$ backgrounds in the above discussion) in the dark matter sector, which in turn would be sensitive to the dark energy modeled by the cosmological
constant $\Lambda$. This leads us naturally to the last topic of this talk - the realization of dark energy in the metastring
approach to quantum gravity.

\section{Quantum gravity, metastrings and dark energy} 

In this section we give a new interpretation~\cite{rBHM8} of dark energy from this novel point of
view of string theory (and quantum gravity) \cite{Freidel:2013zga, Freidel:2014qna, Freidel:2015pka, Freidel:2015uug, Freidel:2016pls, Freidel:2017xsi, Freidel:2017wst, Freidel:2017nhg, Freidel:2018apz, Freidel:2019jor}. 
Ever since the seminal discovery of dark
energy in the late 1990s~\cite{Riess:1998cb}, string theory (viewed as a consistent theory of quantum gravity and matter)
has been attempting to deal
with this central ingredient of fundamental physics.
(For the most recent measurements of the Hubble constant and the associated discrepancies (see~\cite{Riess:2016jrr}). 
The existence of de Sitter space (dS) as a solution in string theory (and dark energy in the observable universe) is still considered an 
outstanding open question~\cite{Danielsson:2018ztv}, and the interest in this fundamental issue has been recently reignited in~\cite{Obied:2018sgi}. %(see also~\cite{Andriot:2019wrs}). 

\subsection{Metastring theory and dark energy}

We  now explain how the generalized geometric formulation of string theory discussed above provides for an effective description of dark energy that is consistent with de Sitter spacetime. 
This is essentially due to the theory's chirally and non-commutatively~%\eqref{e:CnCR1} 
doubled realization of the target space 
and the stringy effective action on the doubled non-commutative~%\eqref{e:CnCR1} 
spacetime $(x^a,\tx^a)$ 
\be
  S_{\text{eff}}^{\textit{nc}}
  =\iint \text{Tr} \sqrt{g(x,\tx)}\, \big[R(x,\tx) +L_m(x,\tx) +\dots\big],
\label{e:ncEH}
\ee
where the ellipses denote higher-order curvature terms induced by string theory. %~\eqref{e:Seff}.  
(Here we have included the matter Lagrangian $L_m$
as well.) %Owing to~\eqref{e:CnCR1}, 
This $S_{\text{eff}}^{nc}$ clearly expands into numerous terms with different powers of $\lambda$, which upon $\tx$-integration and from the $x$-space vantage point produce
various effective terms.
To lowest (zeroth) order of the expansion in the non-commutative parameter $\lambda$
of $S_{\text{eff}}^{\textit{nc}}$ takes the form:
\be
S_d = - \iint\! \sqrt{-g(x)} \sqrt{-\tilde{g}(\tx)} [R(x) + \tilde{R}(\tx)],
\label{e:TsSd}
\ee
a result which first was obtained almost three decades ago,  effectively neglecting $\w_{AB}$ %in~\eqref{e:CnCR} 
by assuming that $[\hat x^a,\htx_b]=0$~\cite{Tseytlin:1990hn}. 
In this leading limit, the $\tx$-integration in the first term %of~\eqref{e:TsSd} 
defines the gravitational constant $G_N$, and in the second term produces a {\it positive} cosmological constant constant $\L>0$. 
In particular, we are lead to the following low energy effective action valid at long distances of the observed accelerated universe (focusing on the relevant $3{+}1$-dimensional 
spacetime $X$, of the ${+}\,{-}\,{-}\,{-}$ signature):
\be
S_{\text{eff}} =\frac{-1}{8 \pi G}\int_X \sqrt{-g}
  \big( \L + {\textstyle\frac{1}2} R + {\cal O}(R^2)
\big),
 \label{e:Seff}
\ee
with $\L$ the positive cosmological constant (corresponding to the scale of $10^{-3}$ eV) and
the ${\cal O}(R^2)$ denote higher order corrections (which are also required by the sigma model of string theory~\cite{rF79b}).

It also follows from this construction that the weakness of gravity is determined by the size of the canonically conjugate dual space, while the smallness of the cosmological constant is given by its curvature. 
(Higher order terms in $\lambda$ produce various forms of dark energy~\cite{Joyce:2014kja}
and this may even provide for a way of addressing the recent conflicting measurements of the Hubble constant~\cite{Riess:2016jrr}.)
Given this action, we may proceed reinterpreting~\cite{Tseytlin:1990hn}: integrate out the dual spacetime coordinates, write
the effective action as
  $\bar{S} \sim \tilde{V} \int_X\!\! \sqrt{-g(x)} R(x)+...,$
where
  $\tilde{V} = \int_{\tilde X}\!\! \sqrt{-\tilde{g}(\tx)},$
and then relate the dual spacetime volume to the observed spacetime volume as
$\tilde{V} \sim V^{-1}$ (T-duality). This produces 
an ``intensive'' effective action
~\cite{Tseytlin:1990hn}
\be
 \bar{S} 
= \frac{\int_X\! \sqrt{-g(x)} \big(R(x) +L_m(x)\big)}{ \int_X\! \sqrt{-g(x)}}+\dots
 \label{e:TsLb}
\ee
By concentrating on the classical description first (we discuss below 
 quantum corrections and the central role of intrinsic non-commutativity in string theory) we get the following Einstein equations~\cite{Tseytlin:1990hn}
\begin{equation}
 R_{ab} - \frac{1}{2} R g_{ab} +T_{ab} + \frac{1}{2} \bar{S}\, g_{ab} =0,\qquad
T_{ab} \overset{\scriptscriptstyle\text{def}}= \frac{\partial L_m}{\partial g^{ab}} - \frac{1}{2} L_m\,g_{ab}.
% \label{e:}
\end{equation}
We emphasize that our reinterpretation of~\cite{Tseytlin:1990hn}
does not follow the original presentation and intention.
In particular, we directly relate the intensive action (54) to the cosmological constant, 
$\bar{S}\sim \Lambda$. 

Note that this new approach to the question of dark energy (viewed as a cosmological constant) in quantum gravity
is realized in certain stringy-cosmic-string-like~\cite{rGSVY} toy models~\cite{rBHM7}, %rBHM1, rBHM5, rBHM6},
which can be viewed as illustrative of a generic non-commutative phase of F-theory~\cite{rFTh}.
In particular, the``see-saw" formula (54) is directly realized in~\cite{rBHM7} %, rBHM1, rBHM5, rBHM6} 
as $M_\L\,{\sim}\,M^2/M_P$,
where $M_\L$ is the dark energy scale, $M_P$ the Planck scale and $M$, and intermediate scale, coming from the matter sector.

\subsection{Dark energy and dark matter from metastring theory}

Note that in general, to lowest (zeroth) order of the expansion in the non-commutative parameter $\lambda$
of $S_{\text{eff}}^{\textit{nc}}$ takes the following form (that also includes the matter sector and its dual) \cite{BHM}:
\be
S_d = - \iint\! \sqrt{{g(x)}{\tilde{g}(\tx)}} [R(x) + \tilde{R}(\tx)+ L_m (A(x, \tilde{x})) + \tilde{L}_{dm}   ( \tilde{A}(x, \tilde{x})) ],
\label{e:TsSd}
\ee
Here the $A$ fields denote the usual Standard Model fields, and the $\tilde{A}$ are their duals, as predicted by the general formulation of 
quantum theory that is sensitive to the minimal length.
Note that after integrating over the dual spacetime, and after taking into account T-duality, the equation (54) now reads
\be
 \bar{S} 
= \frac{\int_X\! \sqrt{-g(x)} \big(R(x) +L_m(x) + \tilde{L}_{dm} (x)\big)}{ \int_X\! \sqrt{-g(x)}}+\dots
 \label{e:TsLb}
\ee
The proposal here is that the dual sector (as already indicated in the previous section) should be interpreted as the dark matter sector,
which is correlated to the visible sector via the dark energy sector, as discussed in \cite{Ho:2010ca}. 
We emphasize the unity of the description of the entire dark sector based on 
the properties of the dual spacetime, as predicted by the generic formulation of string theory (as a quantum theory with a dynamical Born geometry).

\subsection{Dark energy and radiative stability} 
Let us return to the discussion of the dark energy sector.
The above results from the commutative limit are not stable under loop corrections. which has been addressed in the recent work of Kaloper and Padilla (called the sequester mechanism) who also extended these results to loops of arbitrary order, in the effective field theory~\cite{Kaloper:2013zca}.
In that context, the effective field theory expansion has to have 
another global scale, $s$, 
so that the sequestering action is proportional to 
\begin{equation}
 \int_X \sqrt{-g} \Big[\frac{R}{2G} + s^4 L (s^{-2} g^{ab}) + \frac{\Lambda}{G}\Big]
 + \sigma\Big(\frac{\Lambda}{s^4 \mu^4}\Big),
% \label{e:}
\end{equation}
where $L$ denotes the combined Lagrangians for the matter and
dark matter sectors, $\mu$ is a mass scale and $\sigma(\frac{\Lambda}{s^4 \mu^4})$ is a
global interaction that is not integrated over
\cite{Kaloper:2013zca}. 
This can be provided by our set up: Start with bilocal
fields  $\phi(x, \tx)$~\cite{Freidel:2016pls}, and replace the dual labels $\tx$ and also $\lambda$ (in a coarsest approximation) by the global dynamical scale $s \sim \Delta {\tx} \,{\sim}\,\l^2 \Delta {x}^{-1} $.
Also, normal ordering produces $\sigma$.
This is 
an effective realization of the sequester mechanism 
in a  non-commutative phase of string theory. 
Furthermore, the intrinsic non-commutativity of the zero modes $x$ and $\tx$ in $[x, \tx] = i \lambda^2$ %~\eqref{e:CnCR1} 
corrects  the zeroth order results in $\lambda$ in several ways. 
In particular, it is natural to ask whether the non-zero $\omega^{AB}$ in  $[ \hat{\X}^A, \hat{\X}^B] = i \w^{AB}$ %~\eqref{e:CnCR}  
stabilizes the cosmological constant directly on the level of the effective non-commutative action.
 The fully non-commutative analysis is intricate, but 
for conformally flat metrics, $g_{\mu \nu} = \phi^2 \eta_{\mu \nu}$, the action~\eqref{e:ncEH}--\eqref{e:TsSd} produces a non-commutative $\L \phi^4$ theory, which is a natural non-commutative generalization of 
the effective action for conformal metrics  $\int_X (\partial_{\mu} \phi \partial^{\mu} \phi + \frac{\L}{3}  \phi^4)$, with the
non-commutative product depending on $\lambda$\footnote{See also \cite{Polyakov:1993tp}.}.
 Unlike the commutative limit of the theory, the beautiful results of Grosse and Wulkenhaar~\cite{Grosse:2012uv} demonstrate the non-perturbative solvability of the above non-commutative 
$\L \phi^4$ theory, explicitly showing the finite renormalization of $\L$ in terms of the bare coupling.
 At least in this highly simplified, conformal degree limit, non-commutativity thus can afford a small, radiatively and perhaps even non-perturbatively stable cosmological constant for the non-commutative form of the ``doubled'' effective action.

However, non-commutative field theories have both UV and IR scales and 
the effective description is defined by expanding around self-dual fixed points, 
and 
it is organized by keeping  both 
the Wilsonian UV cutoff as well as the IR scale. This clearly meshes nicely with
the UV and IR aspects of the see-saw formula.
Identifying $M_\L$ and $M_P$ as the IR and UV cut-offs, respectively, the double-scale RG flow 
identifies a self-dual fixed point~\cite{Douglas:2001ba}. %, Grosse:2004yu}. 
Given that the phase-space formulation~\cite{Freidel:2013zga, Freidel:2015uug, Freidel:2016pls, 
 Freidel:2017xsi, Freidel:2017wst} is a T-duality covariant description of string theory, this naturally relates $M_P\,{\to}\,M^2/M_P$ under T-duality.
The prediction of our effective stringy \cite{Polchinski:1991ax} cosmic-string-like models~\cite{rBHM7}  
$M_\L\,{\sim}\,M^2/M_P$ then satisfies these conditions, with $M_P\,{\sim}\,\e\,{=}\,1/{\lambda}$ 
the fundamental energy scale corresponding to the fundamental length $\l$, 
which is consistent with observations 
provided $M$ is a \TeV\ scale. 
We emphasize that the usual spacetime discussion 
of string theory is compatible with local effective field theory,
which does not account for the radiative stability of vacuum energy. What we argue is that
this feature of string theory is an artifact of a spacetime description which is not generic.
The generic formulation of string theory is doubled and generalized-geometric, and intrinsically
non-commutative, and it leads to an effective field theory that is sequestered, and thus, to leading order,
to a radiatively stable vacuum energy. (Including further corrections due to intrinsic
non-commutativity.) Only in a singular limit in which one neglects the intrinsic non-commutativity
and works only within a spacetime section of the general doubled description does one find the usual effective 
field theory, with a spacetime interpretation, and the usual questions regarding the existence of
dS background in string theory
\cite{Danielsson:2018ztv} (and the related issues related to holography in the context of asymptotically de Sitter spaces \cite{Balasubramanian:2002zh} as well as
supersymmetry breaking and the existence of 
a small and positive cosmological constant \cite{Witten:1994cga})\footnote{Note also that Starobinsky inflation~\cite{Starobinsky:1980te} may appear as a natural product of the higher order terms in the $\lambda$ expansion that,
after integrating over $\tilde{x}$ can result  in $\int_X \sqrt{-g} (R + a R^2)$, at the next to leading order in $\lambda$. Starobinsky inflation 
beautifully fits the observed data~\cite{Akrami:2018odb}, and is non-supersymmetric ---
which is consistent with the supersymmetry-breaking nature of our discussion.}

\section{Conclusion: Quantum gravity and the real world}

In this talk we have summarized the recent work on 
quantum foundations of string theory and quantum gravity
\cite{Freidel:2013zga, Freidel:2014qna, Freidel:2015pka, Freidel:2015uug, Freidel:2016pls, Freidel:2017xsi, Freidel:2017wst, Freidel:2017nhg, Freidel:2018apz, Freidel:2019jor}. In particular, we have
discussed intrinsic non-commutativity in quantum gravity
related to a new concept of quantum spacetime, called modular spacetime that
also appears as a habitat for metastring theory  and that is deeply rooted in the
foundations of quantum theory (and, especially, in the concept of modular variables
that goes back to the work of Weyl, Schwinger and Aharonov). Note that this
concept stems from a quantization of spacetime, and not from quantization of gravitational
field/metric. Even the 
flat space is quantized according to our approach
to quantum gravity. 
This allows for superposition and entanglement of spacetimes.
Also, this formulation provides for an explicit construction of spacetime quanta
or qubits (the fully ``compactified" bosonic string), and a new non-perturbative
definition of quantum gravity as ``gravitization of the quantum". Such a fully
dynamical (or ``curved") construction of Born geometry can be approached from the
point of view of ``teleparallel gravity" in which one utilizes the 
at (zero-curvature)
connection and crucially introduces non-zero torsion \cite{einstein}.  This viewpoint is natural
for the rigid structure of Born geometry and it allows for ``curving" of T-duality.
In some sense, by going from our new formulation of quantum mechanics in
terms of modular, or quantum spacetime, with hidden but fixed Born geometry,
and its application to quantum field theory, to an explicit formulation of quantum
gravity that involves dynamical Born geometry, as is the case in metastring theory,
we are retracing (in a purely quantum context) the line of development that
led from special relativity (and fixed Minkowski geometry) and its application to
classical relativistic field theory, to general theory of relativity with a dynamical
spacetime geometry\footnote{There is an interesting connection here to quantum logic \cite{qlogic},
``Logic being to quantum theory what geometry is to gravity'' (R. Sorkin). Quantum
gravity then appears as ``warped logic'' (D. Finkelstein), and as ``third relativity'' (J. A. Wheeler);
first relativity being represented as Galilean, special and general relativity and second relativity as quantum theory.}.

The underlying physical principle here is relative locality - different observers probe different spacetimes; these different spacetimes
are sections of a quantum (modular) spacetime - implying, in general, a dynamical momentum space\footnote{
Momentum space geometry is relevant for the recent progress in scattering amplitudes \cite{ArkaniHamed:2012nw}.}.
According to relative locality, quantum mechanics follows from non-locality that is
consistent with causality. In particular, it follows from the existence of fundamental length and fundamental time
that are consistent with Lorentz symmetry. The geometry of quantum non-locality is Born geometry.
In particular, if the relevant physics lives on a spacetime lattice, the full set of quantum numbers involves the lattice and its dual.
This leads to unitary (modular) variables, and modular spacetime (spacetime and its dual being tied via 
intrinsic non-commutativity that involves the fundamental length/time.)
Then it follows that quantum gravity is dynamical Born geometry found in the metastring approach to quantum gravity.
The effective description is given in terms of quantum field theory with intrinsic non-commutativity - with metaparticle excitations.
The renormalization group has both UV and IR scales - leading to a double RG, with direct relevance for the recent discussion of
the infrared limit of QFT \cite{Strominger:2017zoo}.
This new understanding of quantum field theory with manifest quantum non-locality is
relevant both for particle physics and also for condensed matter physics.
In particular, some of the implications for the real world are dark energy as curvature of the dual spacetime \cite{rBHM8}, and dark matter correlated
to visible matter and dark energy \cite{Ho:2010ca} and represented by metaparticles \cite{Freidel:2019jor}.

Finally, we comment on the problem of vacuum selection.
Born reciprocity \cite{born}
demands a symmetric dynamical geometric structures 
in spacetime  and energy momentum space.
Thus matter and spacetime are put on equal footing (which reminiscent of with some intuitions from F-theory). 
The question is whether Born reciprocity can
be used as a criterion for vacuum selection in quantum gravity, which selects ``maximally symmetric solutions''
both in spacetime (de Sitter space) and in the matter sector (Standard Model and its dual, describing the dark matter
sector).
Then the apparent robustness of the ``genetic code'' (masses and couplings) of particle physics 
(see, for example, \cite{Chan:2015eva})
might be the consequence of an attractor mechanism that makes 
the observed cosmology and particle physics ``universal''
(this is similar to what happens in universal biology of
the ``genetic code'' based on horizontal gene transfer \cite{Argyriadis:2019fwb}). 
The idea here is that a universal ``genetic code'' for particle physics (that is, the particle masses and
couplings, as well as cosmological parameters) can be obtained via a horizontal information 
transfer mechanism that apparently leads to a
universal genetic code. 
In the case of universal cosmology, such horizontal information transfer can be provided by 
an interaction between spacetime quanta via the gravitization of the quantum,
leading to a ``maximally symmetric'' attractor solution.
 It would be fascinated to explore this idea more concretely in the future.

{\bf Acknowledgements:}
DM is grateful to L. Freidel, R. G. Leigh, J. Kowalski-Glikman, 
P. Berglund, T. H{\"u}bsch, 
D. Edmonds, T. Takeuchi, D. C. Dai, D. Stojkovic, J. H. Simonetti, M. Kavic and V. Jejjala,
for  recent illuminating collaborations and insightful discussions. 
DM thanks the organizers of the international conference of the Polish Society on Relativity, the Bangkok workshop on High Energy Theory, the Multimessenger Universe conference
at Penn State's Institute for Gravitation and the Cosmos, the 10th Mathematical Physics Meeting in Belgrade, Serbia, the Miami winter conferences, and the conference on quantum gravity phenomenology
in Granada, Spain, for kind hospitality and invitations to present this talk.
The research of DM is supported in part by the US Department of Energy (under grant DE-SC0020262)
and the Julian Schwinger Foundation.

%%%%%%%%%%%%%%%%%%%%%%%%%%%
%: Bibliography
%\footnotesize
%
%\bibliographystyle{utphys}
%\bibliography{RefsHR}
%\end{document}
%
%\clearpage
\begingroup
\frenchspacing\raggedright
\endgroup

\end{document}